\documentclass[preprint, aps,showpacs,nofootinbib, onecolumn]{revtex4-1}

\usepackage{amsmath}
\usepackage{graphicx}
\usepackage{upgreek}
\usepackage{xcolor}

\newcommand{\tb}[1]{\textcolor{red}{#1}}

\newcommand{\eq}[1]{Eq.~(\ref{#1})}
\newcommand{\eqs}[2]{Eqs.~(\ref{#1}) and (\ref{#2})}
\newcommand{\eqto}[2]{Eqs.~(\ref{#1})-(\ref{#2})}

\newcommand{\fig}[1]{Fig.~\ref{#1}}
\newcommand{\subfig}[2]{Fig.~\ref{#1}(#2)}
\newcommand{\subfigs}[3]{Fig.~\ref{#1}(#2) and \ref{#1}(#3)}

\renewcommand{\vec}[1]{{\bf #1}}

\newcommand{\cref}[1]{Ref.~\cite{#1}}
\newcommand{\crefs}[2]{Ref.~\cite{#1} and \cite{#2}}

\newcommand{\bea}{\begin{eqnarray}}
\newcommand{\eea}{\end{eqnarray}}

\newcommand{\fder}[2]{\frac{\updelta #1}{\updelta #2}}
\newcommand{\pder}[2]{\frac{\partial #1}{\partial #2}}

\newcommand{\F}{\mathcal{F}}

\newcommand{\Lp}{L_{\rm p}}

\newcommand{\rhopr}{\rho_{\rm p}^{\rm r}}
\newcommand{\rhopri}[1]{\rho_{{\rm p}, #1}^{\rm r}}

\newcommand{\Vovi}[1]{V_{{\rm ov}, #1}}

\newcommand{\rhoc}{\rho_{\rm c}}
\newcommand{\kt}{k_{\rm B}T}

\renewcommand{\r}[2]{\rho_{{\rm #1}, #2}}
\newcommand{\cone}[2]{c_{{\rm #1}, #2}^{(1)}}

\begin{document}

\title{Phase diagrams for sticky rods in bulk and in a monolayer from a lattice free-energy functional
for anisotropic particles with depletion attractions}
\author{M.~Mortazavifar and M. Oettel}
\email{Email address: martin.oettel@uni-tuebingen.de}
\affiliation{Institut f\"ur Angewandte Physik, Eberhard Karls Universit\"at T\"ubingen, Auf der Morgenstelle 10, D-72076 T\"ubingen, Germany}

\begin{abstract}
  A density functional of fundamental measure type for a lattice model of anisotropic
  particles with hard-core repulsions and effective attractions
  is derived in the spirit of the Asakura-Oosawa model.
  Through polymeric lattice particles of various size and shape, effective attractions of different strength
  and range between the colloids can be generated.
  The functional is applied to the determination of phase diagrams for sticky rods of length
  $L$ in two dimensions, in three dimensions and in a monolayer system on a neutral substrate. In all cases, there is a competition
  between ordering and gas-liquid transitions. In two dimensions, this gives rise to a tricritical point,
  whereas in three dimensions, the isotropic-nematic transition crosses over smoothly to a gas-nematic
  liquid transition. The richest phase behavior is found for the monolayer system.
  For $L = 2$, two stable critical points are found corresponding to a standard gas-liquid 
  transition and a nematic liquid-liquid transition. For $L = 3$, the gas-liquid
  transition becomes metastable. 
\end{abstract}

\pacs{}
\maketitle

\section{Introduction}

Quite often, lattice models are used to investigate general aspects of the statistical mechanics of phase transitions.
Also, a lattice specific model may be constructed as a simplified version of a certain continuum model of interest which
is, in general, harder to study with analytical methods. The textbook example is the lattice gas
of particles whose hard cores occupy one lattice site, respectively, and nearest neighbors 
attract each other with a finite energy $\epsilon$, see, e.g., \cref{BookHuang}. This 
model shows a gas-liquid transition similar
to simple liquids with isotropic, pairwise attractions between atoms and it can be mapped to the Ising model.

Anisotropic  particles with mutual attractions (e.g., rods on cubic lattices) may show ordering transitions such as a nematic transition
which will compete with a gas-liquid transition. This is a quite relevant class of model systems considering the advances
in the preparation of colloidal solutions with well-defined particle anisotropy \cite{Sac11,Bas13}. But also
the phase behavior of molecular systems where anisotropic molecules interact non-covalently (mostly true for organic
molecules) may be understood in terms of such basic lattice models. 
However, while pure hard-core lattice fluids have received some attention, surprisingly few results on attractive lattice rods are available in the literature.

Theoretical studies of lattice hard rods in two dimensions (2D) and three dimensions (3D) were sparked by DiMarzio~\cite{DiMar61}, having approached the problem from the
context of polymer theory.
DiMarzio calculated the number of possible packings of rods---thus evaluated the entropy---through
approximating the probability of inserting a new rod into a system already containing other rods in a mean-field fashion.
DiMarzio's free energy for rods of length $L\times 1 \times 1$ on cubic lattices leads to a strong first-order nematic transition for $L \ge 4$
\cite{Alb71} and for rods of length $L\times 1$ on square lattices to a continuous nematic transition for $L \ge 4$ \cite{Oet16}. 
Furthermore, DiMarzio's free energy is the same as the one from an exact solution on Bethe-like lattices \cite{Cot69,Dhar11}.
Not unusual for mean-field approaches to the nematic transition, the tendency towards ordering is overestimated in comparison
to simulations. In 2D, these show a nematic transition (demixing between $x$- and $y$-oriented rods) 
for $L \ge 7$ \cite{Ghosh07} which is a critical one. In 3D a transition to a nematic state
with negative order parameter (one minority species) for $L=5,6$ and a transition to an ``ordinary'' nematic state (with one majority species)
for $L \ge 7$ \cite{Gschwind17,Raj17}, is found which is very weakly first order.

Attractions have been considered in the literature mainly for the case of sticky rods (attractions proportional to the number
of touching sites between neighboring rods). For $L=1$, this system is the lattice gas where different types of approximations
have become textbook material \cite{BookHill}. The simplest one, the Bragg-Williams approximation, treats the distribution of particles
randomly and leads to a quadratic dependence of the attractive part of the free energy on the particle density.
It is completely equivalent to the van~der~Waals approximation for simple fluids and accordingly displays a gas-liquid transition.
A more sophisticated one, the Bethe-Peierls or chemical approximation, treats the distribution of pairs of next-neighbor sites randomly 
and gives a phase diagram closer to exact results (the Onsager solution in 2D or simulations in 3D) than the Bragg-Williams approximation.  
For $L>1$, the literature focuses on 2D systems (surface adsorption of flat-lying rods) with approaches combining the DiMarzio entropy with the quasi-chemical
approximation \cite{Lon12} or employing simulations \cite{Lon10}. An interesting variant of surface adsorption considers flat-lying and standing molecules
[we will call this a (2+1)D system] 
which has been treated in \cref{Boehm77} using the DiMarzio-Bragg-Williams approximations and in \cref{Kra92} also 
by simulations.   
For 3D systems of attractive rods we have not found results in the literature.

In this paper, we approach the problem of attracting lattice rods somewhat differently, aiming at a free-energy functional
which should be applicable to homogeneous and inhomogeneous situations. Effective attractions between the rods are
induced by fictitious polymer particles in the spirit of the Asakura-Oosawa (AO) model \cite{AO54,Vri76}. 
These polymer particles interact hard 
with the rods but have no interactions among each other (ideal gas). Consequently there is an exclusion volume around each lattice
rod which polymers cannot occupy. Effective attractions between rods arise from overlapping exclusion volumes
around the rods which release free volume to the polymers, increase their entropy, and decrease the free energy of the system.  

From the technical side, we will derive the free-energy functional for such a lattice AO model with methods known from the continuum
\cite{Schm00,Bra03}. Starting from a free-energy functional for a general hard-rod mixture (lattice rods + polymer particles), the functional
is linearized with respect to the polymer species such that the polymer-polymer direct correlation function
is zero (ideal gas).
Variable attractions between the lattice rods can be induced (such as face and edge interactions with variable strength)
through the selection of size, shape and density of the polymers.
For actual analytical and numerical results, we use the Lafuente-Cuesta (LC) hard-rod functional \cite{Laf02,Laf04}, 
derived from fundamental measure theory (FMT) as a starting point and consider next neighbor (sticky) interactions. 
In the bulk, the LC functional is equivalent to DiMarzio's entropy \cite{Oet16,Gschwind17}. For $L=1$ (lattice gas), the AO treatment
is equivalent to the Bragg-Williams approximation (which we call the ``naive mean-field approximation'' \cite{Mor16}) but for 
$L \ge 2$, the AO model accounts for the limited free volume available to the rods and goes beyond it. Phase diagrams for sticky rods in 
2D, 3D and (2+1)D are calculated which show the interplay of ordering (nematic) transitions and liquid-gas transitions. 
However, the topology of the phase diagrams differ, since the nematic transitions are either Ising critical (2D), first order (3D), or continuous with an onset
at zero density [(2+1)D].

The paper is structured as follows. Section~\ref{sec:model} introduces the AO lattice model and presents a general derivation of an 
FMT-AO functional. In Sec.~\ref{sec:LC}, the Lafuente-Cuesta functional is used to derive explicit functionals for sticky rods 
in 2D, 3D and (2+1)D. The resulting bulk phase diagrams are presented. Finally, Sec.~\ref{sec:summary} gives a short summary and an outlook. 

\section{The Model}
\label{sec:model}

\subsection{The lattice model}
\begin{figure*}[!t]
  \centering\includegraphics[width=0.32\textwidth]{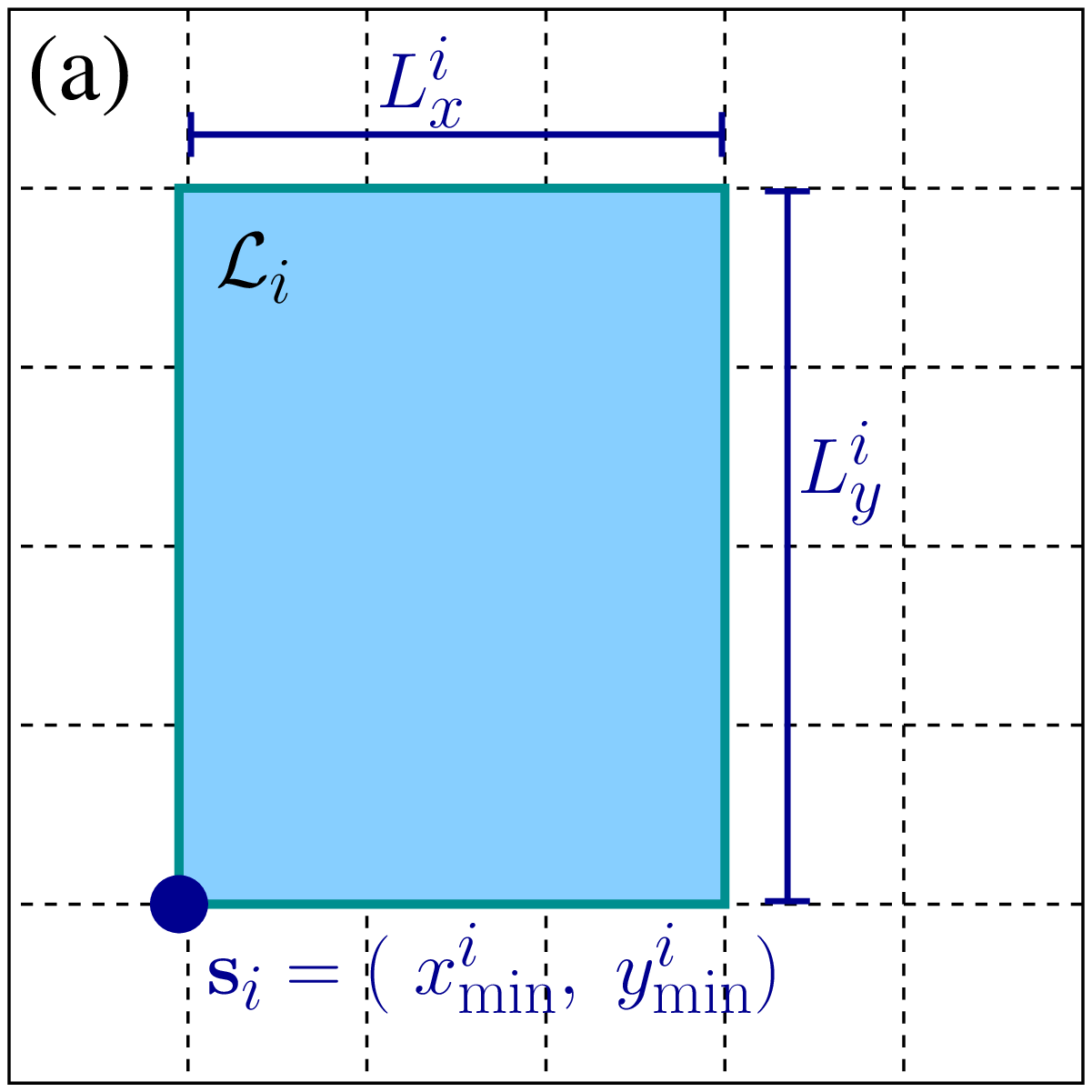}\;\;
  \centering\includegraphics[width=0.32\textwidth]{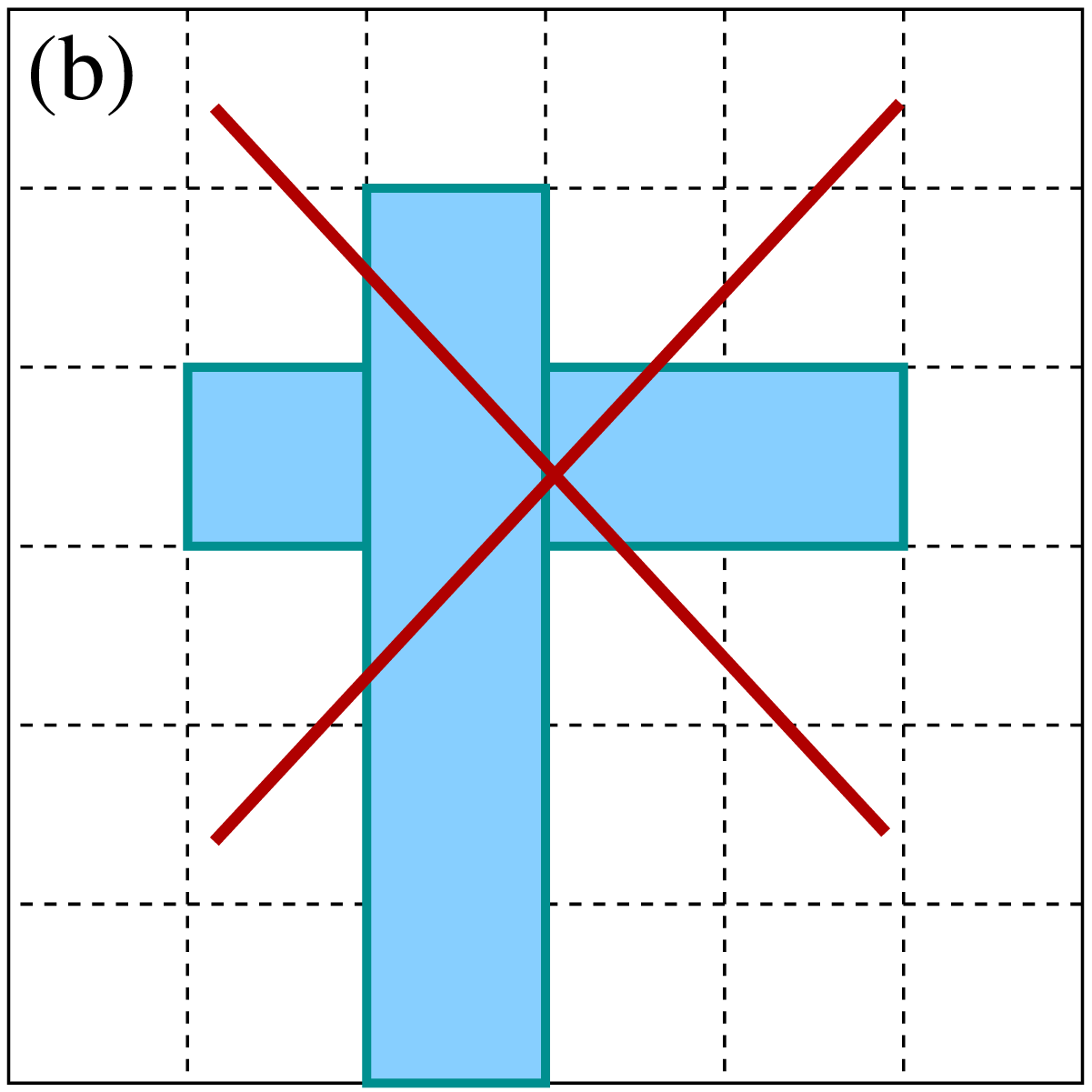}\;\;
  \centering\includegraphics[width=0.32\textwidth]{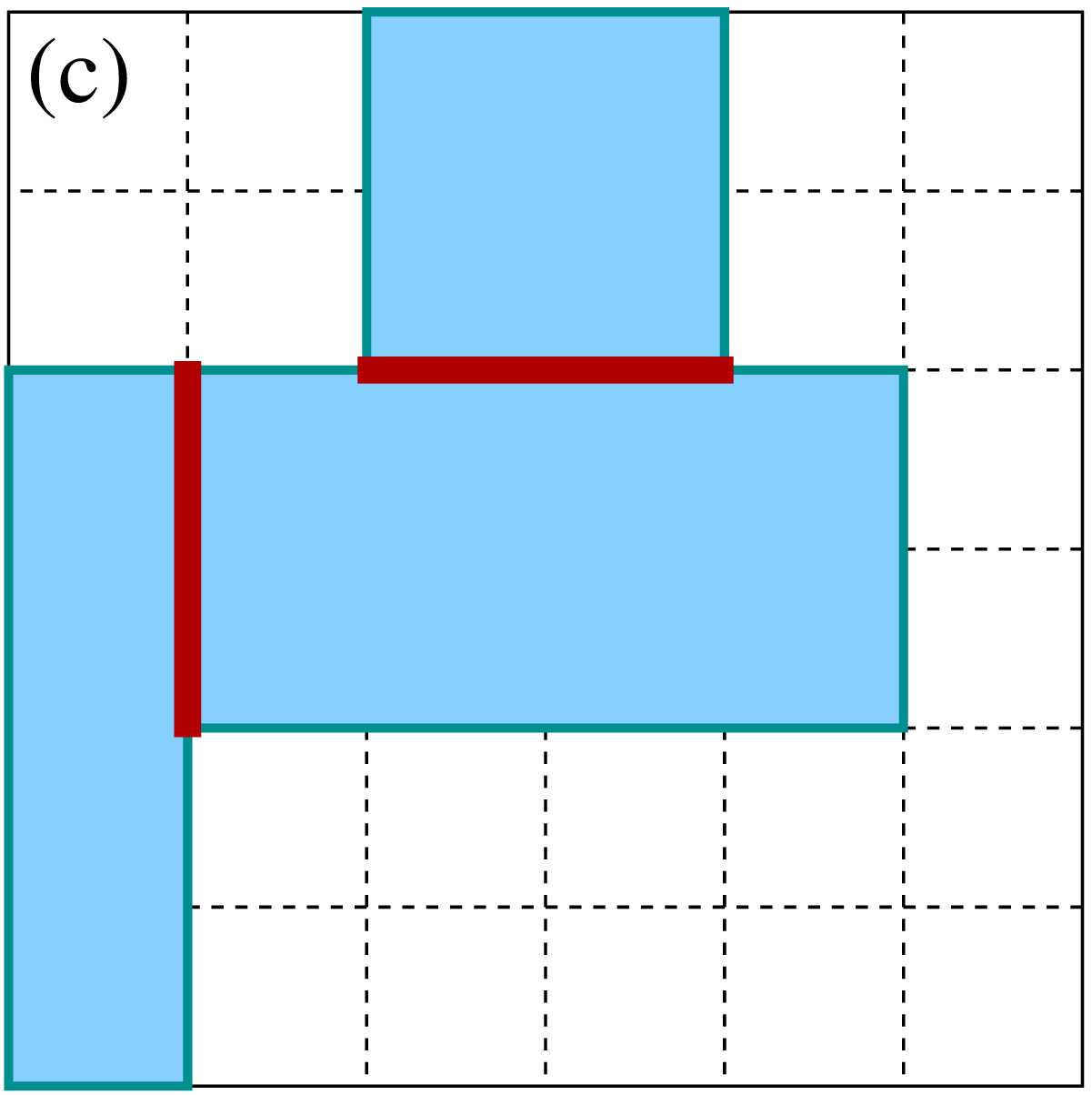}
  \caption{(a) A representation of the lattice model in 2D.
           Particle $\mathcal{L}_i$ is fully specified by its size vector $\vec{L}_i = (L_x^i, L_y^i)$
           and its position (blue dot) at $\vec{s}_i=(x^i_{\rm min}, y^i_{\rm min})$.
           (b) Overlap of two or more particles is forbidden due to their mutual hard-core interaction.
           (c) Short-ranged, sticky attraction between the particles.
           The strength of the attractive interaction is proportional to the number of neighboring lattice sites (length of bold red lines).
           }
  \label{fig:lattice}
\end{figure*}

Consider a simple cubic lattice in $d$ dimensions where
a lattice point $\vec{s}=(s_1, \dots, s_d)$ is specified by a set of $d$ integers $s_i$ [\subfig{fig:lattice}{a}].
The lattice constant sets the unit of length.
The particles of interest are rectangles in 2D and parallelepipeds in 3D.
The state of a particle of species $i$, denoted by $\mathcal{L}_i$,
is fully determined by its position $\vec{s}_i$ and its size vector $\vec{L}_i$.
The size vector $\vec{L}_i = (L_1^i, \dots, L_d^i)$ specifies the extent of the particle along each Cartesian direction.
The position vector $\vec{s}_i$ is given by the corner whose lattice coordinates are minimal each.
The particles are assumed to have entropic interactions as well as energetic attractions.
The entropic interaction prohibits overlapping of two or more particles [\subfig{fig:lattice}{b}].
A pairwise attraction $u_{{\rm att},ij}$ can be expressed as a function of distance $\vec{s}_i-\vec{s}_j$
between particles of species $i$ and $j$. We will consider effective attractions between particles induced by
polymeric depletants of general type, but in actual numerical calculations
we will only consider the limit of sticky rods
where the attractive interaction between two particles is
proportional to the number of their neighboring lattice sites [\subfig{fig:lattice}{c}].

\subsection{Classical density functional theory}
We will employ classical density functional theory (DFT)
to investigate the system.
In classical DFT,
the grand potential functional for a $\nu$-component mixture is a unique functional of the set of one-body density profiles $\{\rho_i(\vec s)\}$
where $i$ is the species index. Densities are computed as number of particles per lattice site.
The equilibrium density profiles  $\rho_i^{\rm eq}$ minimize the grand potential functional \cite{Eva79},
\bea
   \fder{\Omega\left[\{\rho_i\}\right]}{\rho_i}\Bigg|_{\rho_i=\rho_i^{\rm eq}} = 0\;.
\eea
The grand potential functional is the Legendre transform of the total free energy of the system,
i.e., the sum of the intrinsic free-energy functional $\F$ and the interaction energy of each species with an external potential 
$V_{i}^{\rm ext}$.
For a lattice model the grand potential reads,
\bea
  \Omega\left[\{\rho_i\}\right] &=& \F\left[\{\rho_i\}\right] 
 + \sum_{i=1}^\nu \sum_\vec{s} \rho_i\left(\vec{s}\right) V_{i}^{\rm ext} \left(\vec{s}\right)
 - \sum_{i=1}^\nu \mu_{i} \sum_\vec{s} \rho_i\left(\vec{s}\right) \;,
\eea
where $\mu_i$ is the chemical potential of species $i$ and
the integrals appearing in the continuum become sums over discrete lattice positions $\vec{s}$.
The intrinsic free-energy functional is further decomposed into an ideal gas contribution $\F^{\rm id}$,
\bea
  \beta\F^{\rm id}\left[\{\rho_i\}\right] &=& \sum_{i=1}^\nu \sum_\vec{s} \beta f^{\rm id}\left(\rho_i\left(\vec{s}\right)\right)\;,
 \quad
 {\rm with}\; \beta f^{\rm id}(\rho) = \rho \left(\log\left(\rho\right)-1\right)\;.
  \label{eq:dft_ideal}
\eea
and an excess (over ideal) part $\F^{\rm ex}$ due to the interaction between the particles.
Here, $\beta=1/(\kt)$ is the inverse temperature.
For the following, we need to assume that we know an excess functional for a multicomponent
system of hard particles. In the continuum, fundamental measure theory (FMT) provides
such functionals (for a review see \cref{Rot10}), and a lattice extension to multicomponent hard rods
has been derived by Lafuente and Cuesta \cite{Laf02,Laf04}.
In the framework of FMT, the excess free-energy density $\Phi=\beta f^{\rm ex}$
is expressed as a function of a set of weighted (smeared-out) densities $n^{\vec \alpha}$.
Each weighted density is computed as the sum over species of convolutions of a corresponding weight function $w_i^{\vec \alpha}$
and the density profile $\rho_i$.
For the lattice model, we assume $\F^{\rm ex}$ can be expressed in the following FMT form:
\bea
   \beta \F^{\rm ex}\left[\{\rho_i\}\right] &=& \sum_\vec{s} \Phi\left(\{n^\vec{\alpha}\}\right)\;, \nonumber \\
   {\rm with}\; n^\vec{\alpha}\left(\vec{s}\right)
                   &=& \sum_{i=1}^\nu \sum_{\vec{s}'} \rho_i\left(\vec{s}'\right) w^\vec{\alpha}_i\left(\vec{s}\!-\!\vec{s}'\right)
                    =  \sum_{i=1}^\nu\!\left(\rho_i \ast w^\vec{\alpha}_i\right)(\vec{s})\;,
   \label{eq:dft_fex_general}
\eea
where $\ast$ denotes the discrete convolution.

\subsection{The AO model}
\begin{figure*}[!t]
  \centering\includegraphics[width=0.32\textwidth]{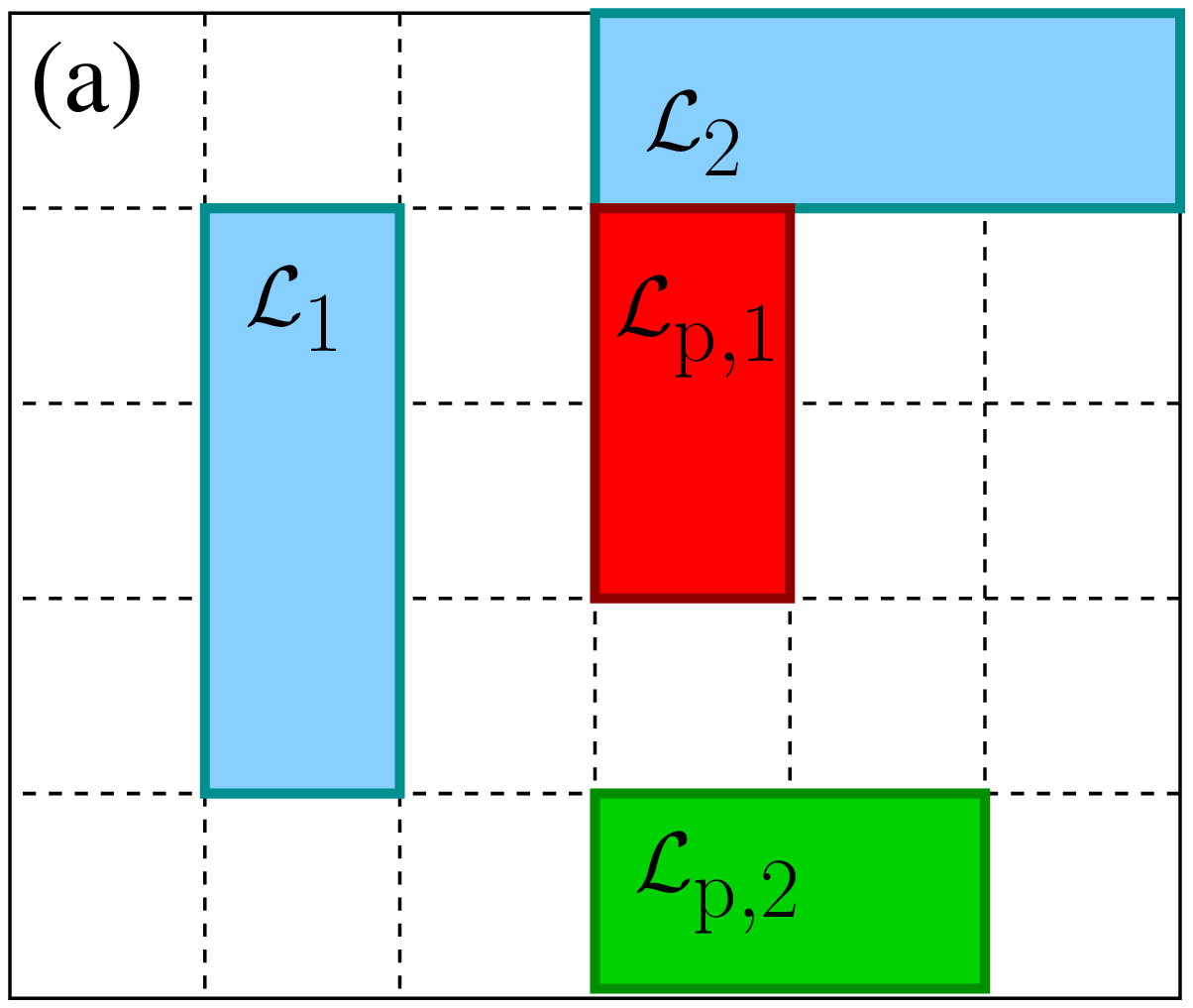}\;
  \centering\includegraphics[width=0.32\textwidth]{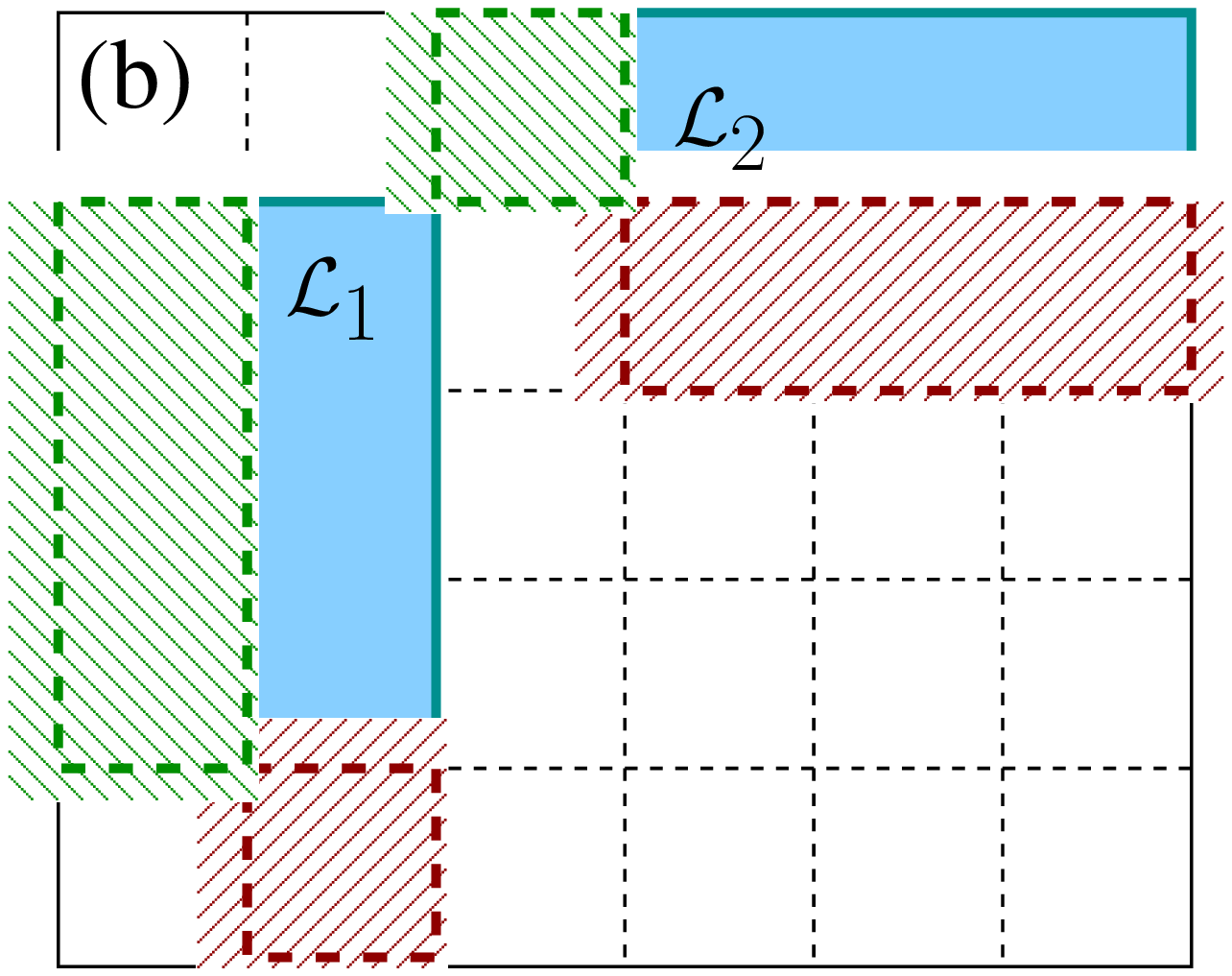}\;
  \centering\includegraphics[width=0.32\textwidth]{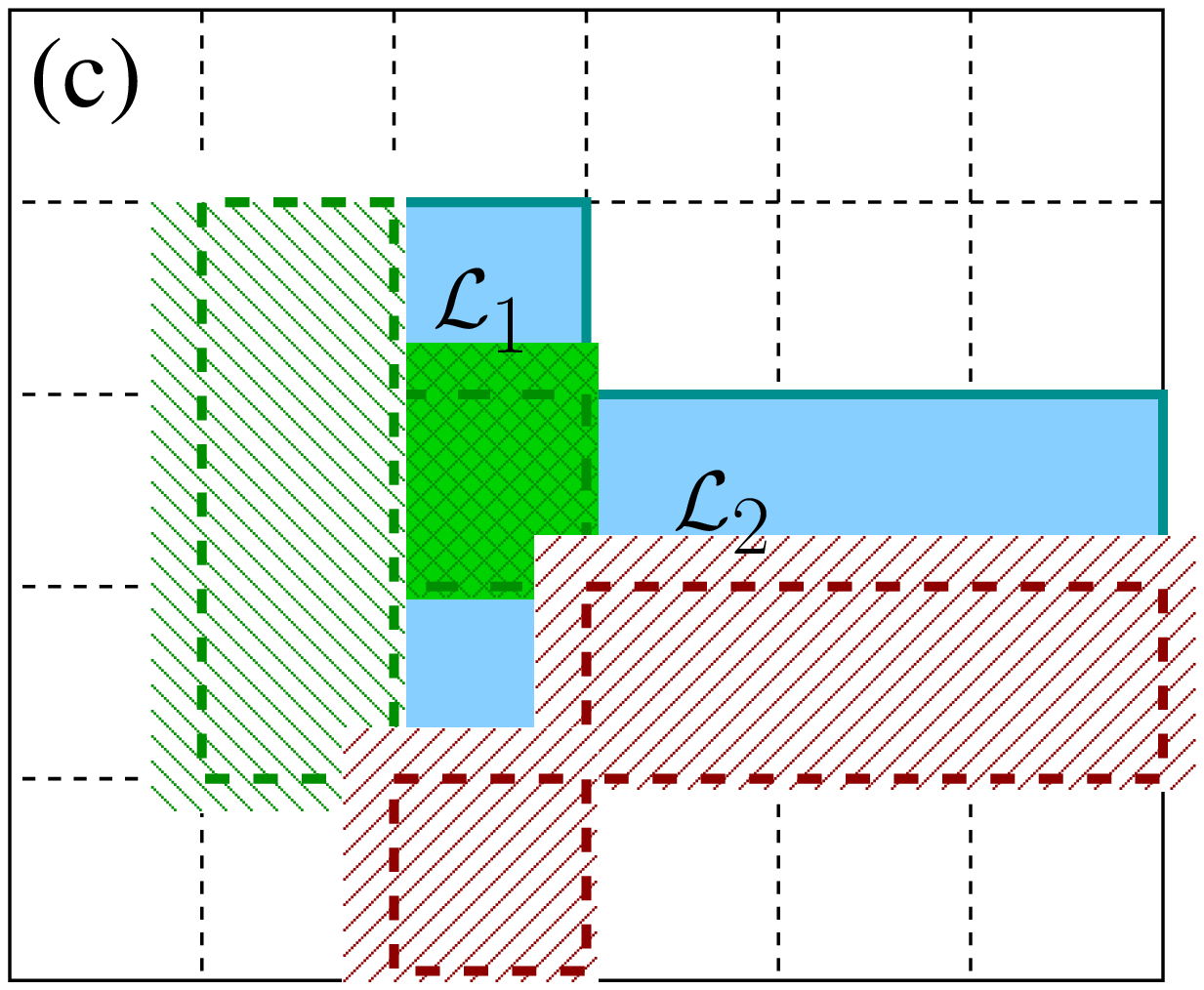}
  \centering\includegraphics[width=0.32\textwidth]{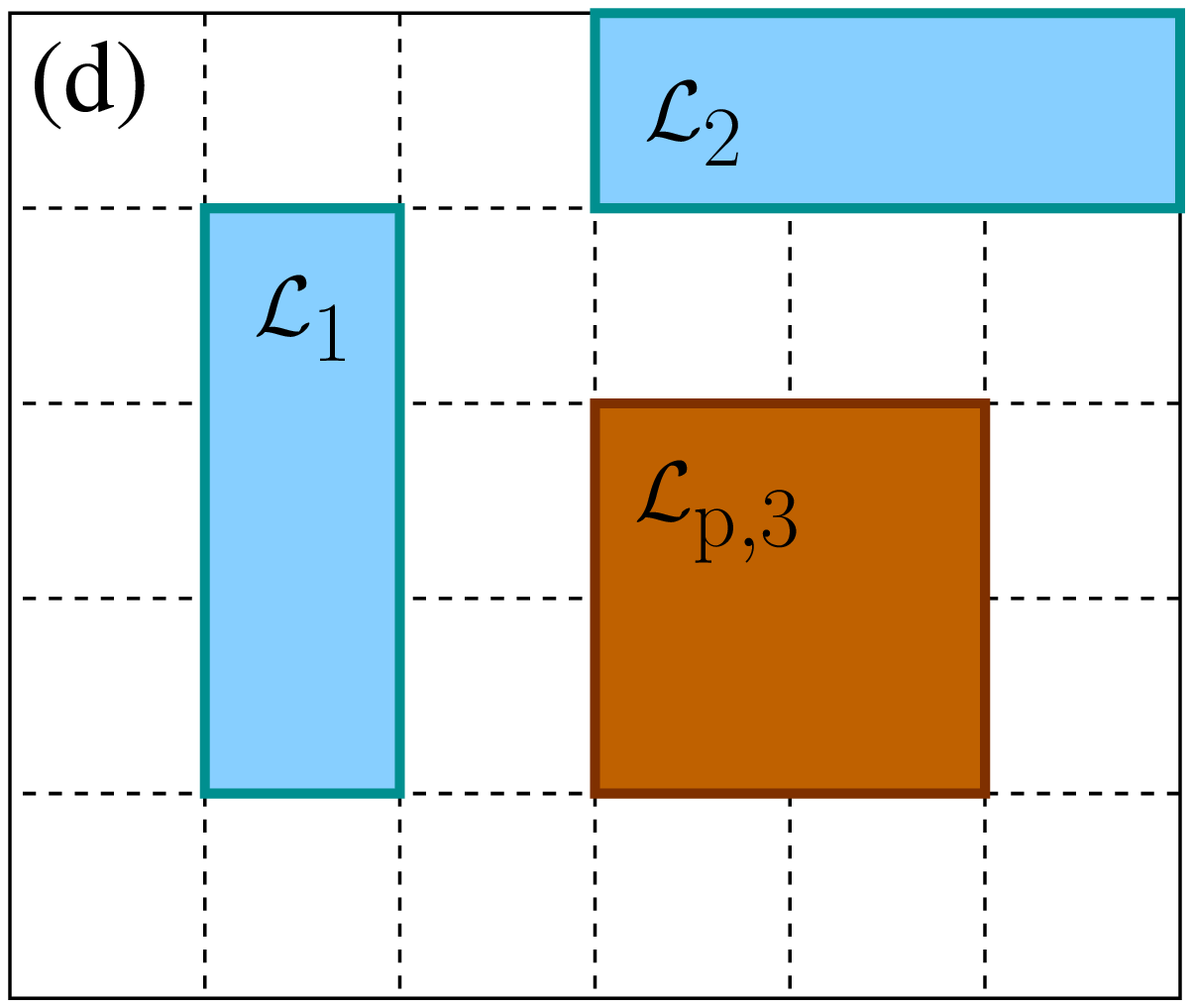}\;
  \centering\includegraphics[width=0.32\textwidth]{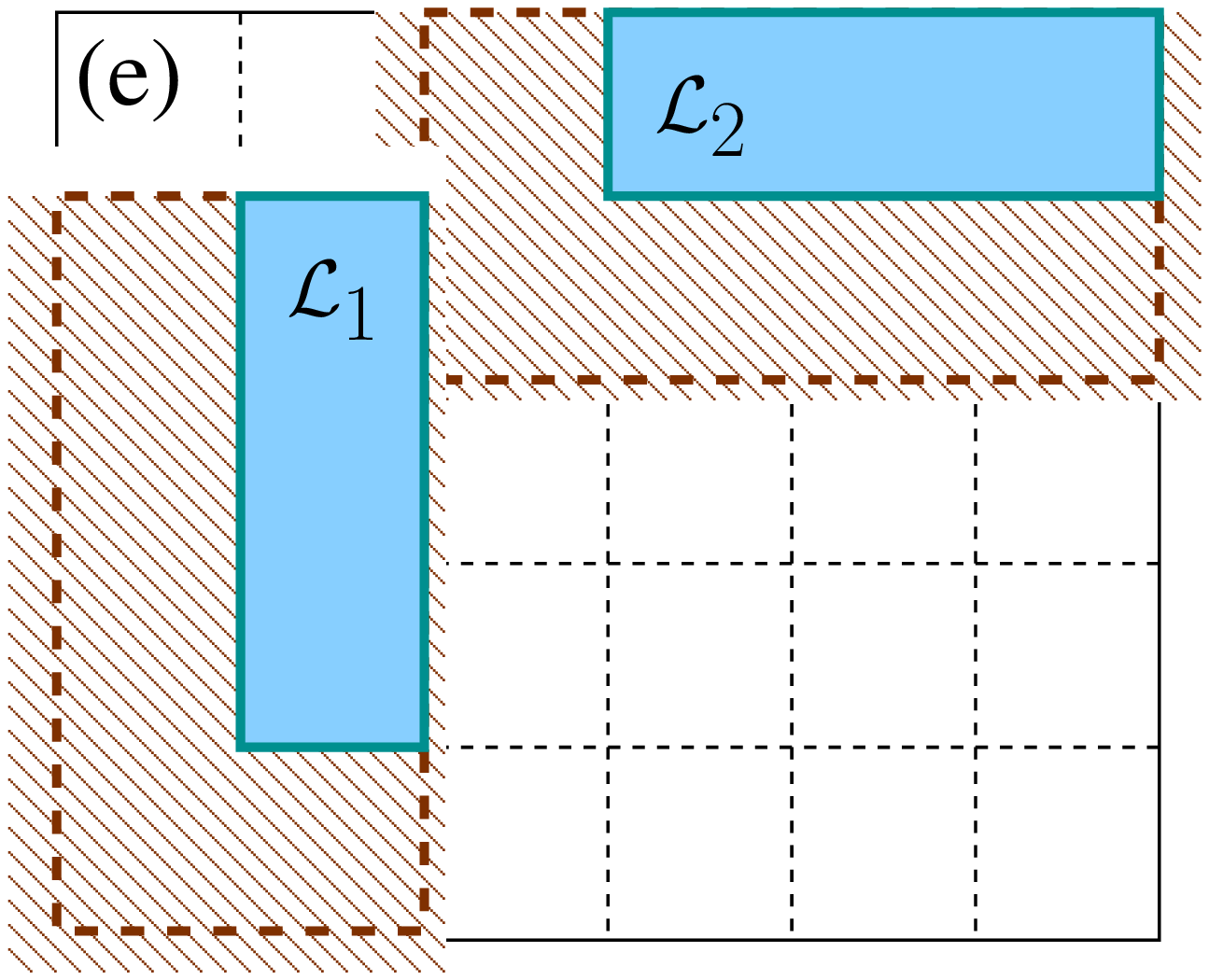}\;
  \centering\includegraphics[width=0.32\textwidth]{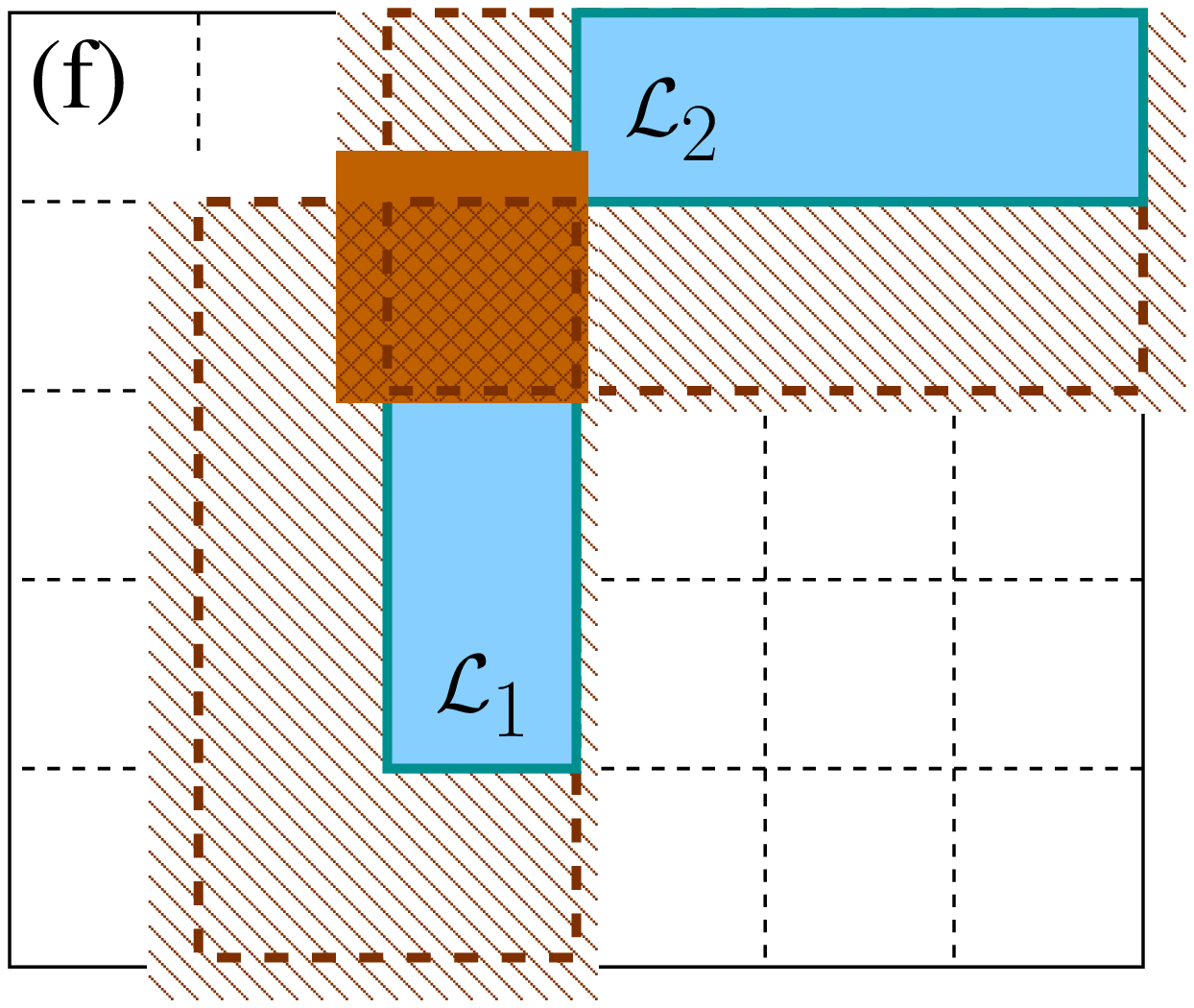}
  \caption{A representation of a mixture of colloidal rods and polymers in the lattice model in 2D.
           Since a polymeric rod occupying only one lattice point does not induce a depletion layer,
           the minimum allowed polymer size is $\Lp=2$.
           Here we have shown two types of such polymers, $\mathcal{L}_{{\rm p},1}$ and $\mathcal{L}_{{\rm p},2}$  with $\vec{L}_{{\rm p}, 1}=(1,2)$ and $\vec{L}_{{\rm p}, 2}=(2,1)$ in (a),
           and $\mathcal{L}_{{\rm p},3}$ with $\vec{L}_{{\rm p},3}=(2,2)$ (d).
           Due to our convention on specifying the position of a particle, the depletion layers
           around the blue colloidal rods are asymmetric as shown in (b) and (e).
           Overlap of the excluded volumes corresponding to one polymeric rod species increases their available free volume and results in an effective attraction between colloidal rods.
           The effective attraction is proportional to the overlapping area as well as the corresponding polymer reservoir density $\rhopri{j}$.
           The anisotropic polymeric rods, i.e, $\mathcal{L}_{{\rm p},1}$ and $\mathcal{L}_{{\rm p},2}$, induce solely sticky attractions 
           along the rod axis as shown in (c) (the corresponding overlap area is shown by the dark green area),
           while $\mathcal{L}_{{\rm p},3}$ induces an edge-edge attraction as well (shown in (f) by the corresponding 
           overlap area in dark brown).
           }
  \label{fig:ao}
\end{figure*}

The short-ranged attractions between the lattice particles are induced by depletion interactions as in the AO model \cite{AO54, Vri76}.
Consider non-adsorbing polymeric particles which do not have any mutual interaction,
but a hard-core interaction with the particles in the system (we refer to the latter as colloidal particles).
Due to the hard-core interaction of the polymeric and colloidal particles,
there exists an excluded volume enclosing the colloidal particles 
which is composed of the proper volume of a colloidal particle itself together with a depletion layer and which the polymers are not allowed to enter.
Note that  polymeric particles occupying only one lattice site (i.e., with size $\Lp=1$ in all lattice directions) do not induce 
an extra depletion layer.
Hence a minimal polymeric particle inducing attractions is a rod with length $\Lp=2$ in one direction and length 1 in the other directions.
Such a polymer species will induce depletion attractions only along the direction where its length is 2. Hence one needs additional
polymer rod species with length greater than one in the other lattice directions to induce corresponding depletion attractions.
Due to our convention of specifying the position of a particle,
the depletion layers are asymmetric in the lattice model (see \fig{fig:ao}).
Overlap of the excluded volumes (corresponding to polymer species $j$) of two colloidal particles of species $i$ and $i'$
increases the available free volume for them, hence their entropy.
This results in an effective attraction between colloidal particles $u_{j, \rm att}^{ii'}$ associated with polymer species $j$.
The induced effective attraction $u_{j, \rm att}^{ii'}$
is proportional to the overlap volume $\Vovi{j}^{ii'}$ of the corresponding excluded volumes,
as well as to the osmotic pressure of polymer species $j$.
Assuming that the system is coupled to reservoirs of polymers which sustain
the chemical potential of each polymer species in the system at a constant value $\mu_{{\rm p}, j}$,
the osmotic pressure of each polymeric rods is equivalent to its corresponding reservoir polymer density $\rhopri{j}$.
In summary,
\bea
   \beta u_{j, \rm att}^{ii'} &=& -\rhopri{j}\;\Vovi{j}^{ii'}\;\quad{\rm with}\;\rhopri{j}=e^{\beta\mu_{{\rm p}, j}}\;.
   \label{eq:ao_summary}
\eea
One sees that the reservoir density $\rhopri{j}$ is equivalent to an inverse temperature.
Note that $\Vovi{j}^{ii'}$, and, consequently, the range of attraction, is determined by the size vector of polymer species $j$ (see \fig{fig:ao}).
As the 2D example of \fig{fig:ao} illustrates, for polymeric rods of length $(2, 1)$ there is only a nonzero overlap
of colloidal excluded volumes if the colloidal rods touch each other along the $x$ direction, and the overlap volume (overlap area in 2D) is given by the number of 
touching sites. Hence these rods induce sticky interactions along the $x$ direction [\subfig{fig:ao}{c}]. Likewise, polymeric rods of length $(1,2)$
induce sticky interactions along the $y$  direction. Polymeric ``squares'' of length $(2,2)$ additionally introduce
edge-edge attractions between the colloidal rods [\subfig{fig:ao}{f}].

\subsection{An FMT functional for short-ranged attractions}
In order to construct an FMT functional for the AO model,
consider a mixture of $n$ species of colloidal rods $\mathcal{L}_{{\rm c},i}$
and $m$ species of non-adsorbing polymers $\mathcal{L}_{{\rm p}, j}$
whose density profiles are denoted by $\rho_{{\rm c}, i}$ and $\rho_{{\rm p}, j}$, respectively.
We start from the excess free-energy density   of 
an $(n+m)$-component mixture of hard  rods (HR) $\Phi^{(n+m)}_{\rm HR}$.
The polymeric rods are not interacting with each other,
and therefore we will assume that
for all combinations of polymer species $j$ and $j'$,
their corresponding second-order direct correlation function $c^{(2)}_{{\rm pp}, jj'}$ vanishes.
Since the direct correlation function is defined by
\bea
 c^{(2)}_{{\rm pp}, jj'}\left(\vec{s} - \vec{s}'\right) &=& -\fder{^2 \beta \F_{\rm ex}^{\rm AO}}{\r{p}{j}\left(\vec{s}\right)\updelta\r{p}{j'}\left(\vec{s}'\right)}\;,
\eea
the terms in the excess free-energy density
should be either constant or linear in polymer densities \cite{Schm00}.
Hence the excess free-energy density of the AO model can be obtained
by linearizing $\Phi^{(n+m)}_{\rm HR}$ with respect to polymer densities $\r{p}{j}(\vec s)$.
Since $\Phi^{(n+m)}_{\rm HR}$ depends on the polymer densities only through the weighted densities
$n_{\rm p}^\alpha = \sum_j \r{p}{j} \ast w_{{\rm p},j}^\alpha$, this results in the free-energy density
\bea
   \Phi_{\rm AO}\left( \{n_{\rm c}^\alpha(\vec s),n_{\rm p}^\alpha(\vec s) \}\right) &=& 
   \Phi^{(n)}_{\rm HR}\left(\{n^\alpha_{\rm c}(\vec s)\}\right) + 
     \sum_\alpha \frac{\partial\Phi^{(n+m)}_{\rm HR}}{\partial n_{{\rm p}}^\alpha(\vec s)}\Bigg|_{\substack{\rho_{{\rm p}, j} = 0\\j=1\dots m}}
   n^{\alpha}_{\rm p}(\vec s)
 \label{eq:AO_Philin}
\eea
where $n_{\rm c}^\alpha(\vec s)$ and $n_{\rm p}^\alpha(\vec s)$ denote weighted densities for colloids and polymers, 
respectively. Likewise, $w_{{\rm c},i}^\alpha(\vec s)$ and $w_{{\rm p},j}^\alpha(\vec s)$ are weight functions
for colloids of species $i$ and polymers of species $j$. 
Note that the derivative has to be evaluated at zero polymer density, and thus it will depend only on
$n_{\rm c}^\alpha(\vec s)$.
Equivalently, the excess free-energy density $\Phi_{\rm AO}=\beta f_{\rm AO}^{\rm ex}$ can be written as
\bea
 \Phi_{\rm AO}\left(\vec s\right) &=& \Phi^{(n)}_{\rm HR}\left(\{n^\alpha_{\rm c}(\vec s)\}\right) - 
   \sum_{j=1}^m \cone{p}{j}(\vec s) \; \rho_{{\rm p}, j}(\vec s)\;,
 \label{eq:AO_lin}
\eea
where we have introduced $\cone{p}{j}$, the first-order direct correlation function for polymer species $j$
defined as
\bea
  \cone{p}{j}(\vec s) &=&
   - \fder{\beta \F_{\rm ex}^{\rm AO}}{\r{p}{j}(\vec s)}\Bigg|_{\substack{\rho_{{\rm p}, j'} = 0\\j'=1\dots m}}
  \nonumber\\&=& -\sum_\alpha\sum_{\vec s'}\left\{
  \frac{\partial\Phi^{(n+m)}_{\rm HR}}{\partial n_{{\rm p}}^\alpha(\vec s')}\Bigg|_{\substack{\rho_{{\rm p}, j'} = 0\\j'=1\dots m}}\; w_{{\rm p}, j}^\alpha(\vec s' - \vec s)
\right\}\nonumber\\
 &=&-\sum_\alpha\left\{\left(
  \frac{\partial\Phi^{(n+m)}_{\rm HR}}{\partial n_{{\rm p}}^\alpha}\Bigg|_{\substack{\rho_{{\rm p}, j'} = 0\\j'=1\dots m}}\;\hat\ast\; w_{{\rm p}, j}^\alpha
\right)( \vec s)\right\}
    \;.
 \label{eq:AO_conep}
\eea
Here $\hat\ast$ is a modified convolution operator.
Therefore
the total free-energy functional for the AO model can be written as follows:
\bea
\label{eq:ao_F}
   \F_{\rm AO}\left[\{\rho_{\rm c}\}, \{\rho_{\rm p}\}  \right] 
   &=& \F^{\rm id}_{\rm c}\left[\{\rho_{\rm c}\}  \right]+
       \F^{\rm id}_{\rm p}\left[\{\rho_{\rm p}\}  \right]
      + \F^{\rm ex}_{\rm AO}\left[\{\rho_{\rm c}\}, \{\rho_{\rm p}\}  \right]\;,\\
    \beta\F^{\rm ex}_{\rm AO}\left[\{\rho_{\rm c}\}, \{\rho_{\rm p}\}  \right]
  &=& \sum_\vec{s} \Phi_{\rm AO}\left(  \vec s \right)\;,
  \label{eq:ao_Fex}
\eea
where the ideal gas part is given by  \eq{eq:dft_ideal} and the excess part by \eq{eq:AO_lin}.
Since we are dealing with a semi-grand ensemble, where the number of colloidal particles
and the chemical potential of polymers is conserved,
a more appropriate quantity for minimization is the semi-grand free energy $\F'_{\rm AO}$
which is the Legendre transformation of $\F_{\rm AO}$
with respect to polymer densities:
\bea
   \F'_{\rm AO}\left[\{\rho_{\rm c}\}, \{\rho_{\rm p}\}  \right] 
   &=& \F_{\rm AO}\left[\{\rho_{\rm c}\}, \{\rho_{\rm p}\}  \right]
       - \sum_{j=1}^m \mu_{{\rm p},j}\;\sum_\vec{s} \rho_{{\rm p},j}(\vec{s})
  \label{eq:ao_Fp}
\eea
For a given colloidal density profile $\{\rhoc\}$, the equilibrium density of each polymer species is obtained
by minimizing $\beta\F'$ with respect to the corresponding polymer density.
\bea
  \fder{\beta\F'}{\rho_{{\rm p}, j}}\Bigg|_{\rho_{{\rm p}, j}(\vec s)=\rho_{{\rm p}, j}^{\rm eq}} = 0 \Rightarrow
  \rho_{{\rm p}, j}^{\rm eq}(\vec{s}) = \rhopri{j}\;e^{ \cone{p}{j}(\vec{s})}\;,
  \label{eq:ao_rpeq}
\eea
where $\rhopri{j}$ is the reservoir density of polymer species $j$ [see \eq{eq:ao_summary}]
and  $\cone{p}{j}(\vec s)$ is its corresponding first-order direct correlation function defined in \eq{eq:AO_lin}.
As a result, once the polymer reservoir densities $\rhopri{j}$ are specified,
the density profile of polymers is given by an explicit functional of only colloidal densities.
When evaluated in the bulk (constant $\rho_{{\rm c},i}$), $\exp(\cone{p}{j})$ is equivalent to the relative part  of the total volume 
available for polymer species $j$ (free volume fraction) \cite{Mor16}. 
Finally, it is desirable to obtain an effective free energy for  colloidal particles $\F^{\rm eff}_{\rm AO}$ 
which retains only the effect of the depletion interactions induced by the polymers.   
This is achieved by subtracting from $\F'$ those terms which are linear in the polymer densities and at most
linear in the colloid densities \cite{Mor16}.
These subtracted terms are equivalent to the grand potential of the polymers which interact at most with one colloidal particle
and hence do not contribute to the effective attractions. Hence,
\bea
   \beta\F^{\rm eff}_{\rm AO}\left[\{\rho_{\rm c}\}; \{\rhopr\}  \right] 
   &=& \beta\F'_{\rm AO}\left[\{\rho_{\rm c}\}, \{\rho_{\rm p}^{\rm eq}\}  \right]
  \nonumber\\&&
   - \sum_{j=1}^m \rhopri{j}\sum_{\vec s} \left\{
   -1 + \sum_{i=1}^n \sum_{\vec{s}'} \rho_{{\rm c},i}\left(\vec{s}'\right)\left( - f_{ij}(\vec s - \vec s')\vphantom{1^1}\right)
   \right\}\;,
   \label{eq:ao_feff}
\eea
where $f_{ij}\left(\vec{s}-\vec{s}'\right)$ is the Mayor-{\it f} bond for the hard interaction
between a colloidal rod of species $i$ at position $\vec{s}'$
and a polymer of species $j$ at position $\vec{s}$.
The connection to the excluded volume $V_{ij}^{\rm excl}$
for polymer species $j$ around a single colloidal particle of species $i$ (useful later)
is given by
\bea
   V_{ij}^{\rm excl} &=& \sum_{\vec{s}'} \left( - f_{ij}(\vec s')\vphantom{1^1}\right)
       = \prod_{k=1}^d \left( L_k^{{\rm c},i} + L_k^{{\rm p},j}-1\right)\;,
   \label{eq:ao_vexcl}
\eea
where $L_k^{{\rm c},i}$ and $L_k^{{\rm p},j}$ are respectively the $k$th component of
the size vectors of the colloid species $i$ and the polymer species $j$.

The part in $\F^{\rm eff}_{\rm AO}$ resulting from the depletion attractions is given by
subtracting the ideal and excess free energy of the hard rods,
$\beta\F^{\rm eff}_{\rm AO,att}=\beta\F^{\rm eff}_{\rm AO}-\F^{\rm id}_{\rm c}\left[\{\rho_{\rm c}\}\right]- \Phi^{(n)}_{\rm HR}\left(\{n^\alpha_{\rm c}\}\right)$.
In the limit of small colloid densities it is given by
\bea
 \beta\F^{\rm eff}_{\rm AO,att} \approx  \frac{1}{2} \sum_{j=1}^m \sum_{i,i'=1}^n \sum_{\vec s, \vec s'} 
    \rho_{{\rm c},i}(\vec s) \rho_{{\rm c},i'}(\vec s') \; \beta u_{j, \rm att}^{ii'}(\vec s - \vec s') \;.
 \label{eq:naiveMF}
\eea
We call this the {\em naive mean-field approximation}. It sums over all two-particle depletion interactions induced by
polymers. The depletion potential $u_{j, \rm att}^{ii'}(\vec s - \vec s')$ between a colloidal pair of particles belonging to species $i,i'$ 
induced by polymer species $j$ is given by \eq{eq:ao_summary}.
Note that one can in principle
work with negative reservoir densities such that the depletion potential becomes repulsive. For bulk systems (all colloidal
densities are constant), \eq{eq:naiveMF} corresponds to the Bragg-Williams approximation.   

Let us make a few remarks on the possible merits and limitations of the effective free-energy functional derived here. 
$(i)$ The effective AO functional contains multibody attractions if triple
or higher overlaps of excluded volumes around colloidal particles are possible. Below, however, we will present
results on rods with effectively pairwise sticky interactions to demonstrate the use
of the AO model to treat short-ranged, pairwise interactions. 
The treatment of two-particle attractions is usually very difficult in DFT and therefore practical
approximations often resort to a naive mean-field approximation of the type as in \eq{eq:naiveMF}. 
(It works better than one might expect, see the discussion in \cref{Arch17}.)
As is shown,
the AO functional goes beyond it. 
$(ii)$ The density expansion of the AO free energy [\eq{eq:ao_Fp} or \eq{eq:ao_feff}] contains only terms up to
linear order in $\rhopri{j}$, as a result of assuming vanishing direct correlation functions between the polymers.
In the full AO model (with ideal polymers), the polymer-polymer direct correlation function
does not vanish; its virial expansion starts with terms quadratic in the colloid density. Consequently, the
density expansion of the full AO model contains higher-than-linear terms in  $\rhopri{j}$ 
(see \cref{San15} for a corresponding calculation of
virial coefficients in the standard, continuum AO model).
In \cref{BraderThesis} it is extensively argued why the linearization of FMT-based functionals is still a good approximation.
However, we expect deviations for high $\rhopri{j}$
(equivalent to low temperatures). Further improvements in this direction might explore the ideas of 
\cref{Schmidt05} to treat the polymers as clusters in the construction of the functional.
$(iii)$ If an approximation for $\Phi^{(n+m)}_{\rm HR}$ is used in deriving an explicit form
for the AO functional, the low-density expansion of the effective attractive free energy will result in
the correct form of \eq{eq:naiveMF} only if the functional expansion of $\Phi^{(n+m)}_{\rm HR}$ in densities  is correct up to third order.

In the following, as an exemplary case,
we will use the method of  Lafuente and Cuesta \cite{Laf02, Laf04}
for constructing an FMT free-energy functional in the form of \eq{eq:dft_fex_general} for hard rods in the lattice model
and the explicit derivation of an AO functional.
The functional is applied to the case of sticky rods whose long axis is of length $L$ and all other axes are
of length 1 for 2D and 3D systems,
as well as a (2+1)D system, i.e., rods confined to a substrate.
In this paper, we will discuss only the phase diagrams and leave considerations of inhomogeneous systems to later work.
For this purpose, in each case an expression for the effective free-energy density
is presented and the necessary equilibrium properties are obtained.

\section{FMT-AO from the Lafuente-Cuesta functional}
\label{sec:LC}

\begin{figure*}[!t]
  \centering\includegraphics[width=0.23\textwidth]{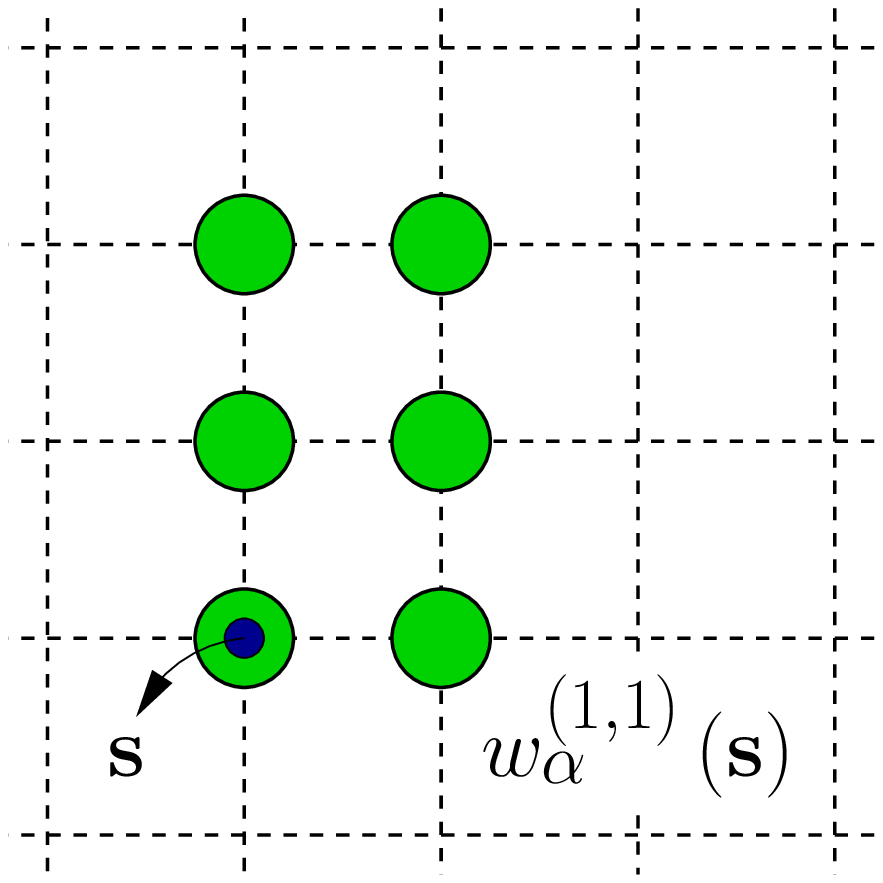}\;\;
  \centering\includegraphics[width=0.23\textwidth]{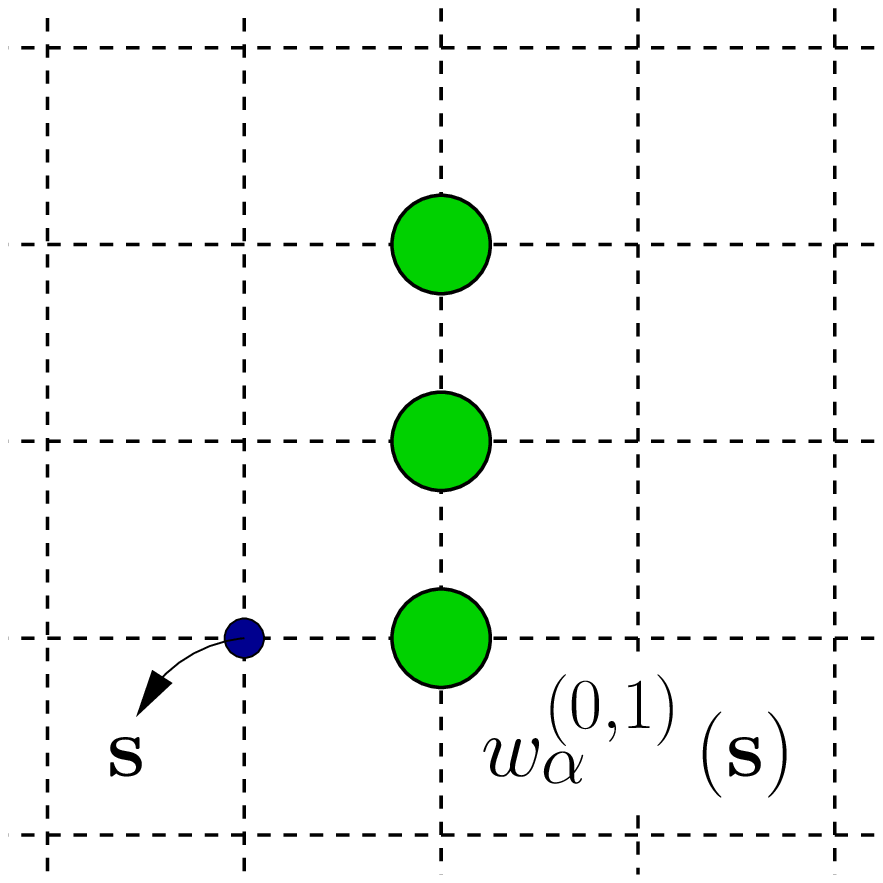}\;\;
  \centering\includegraphics[width=0.23\textwidth]{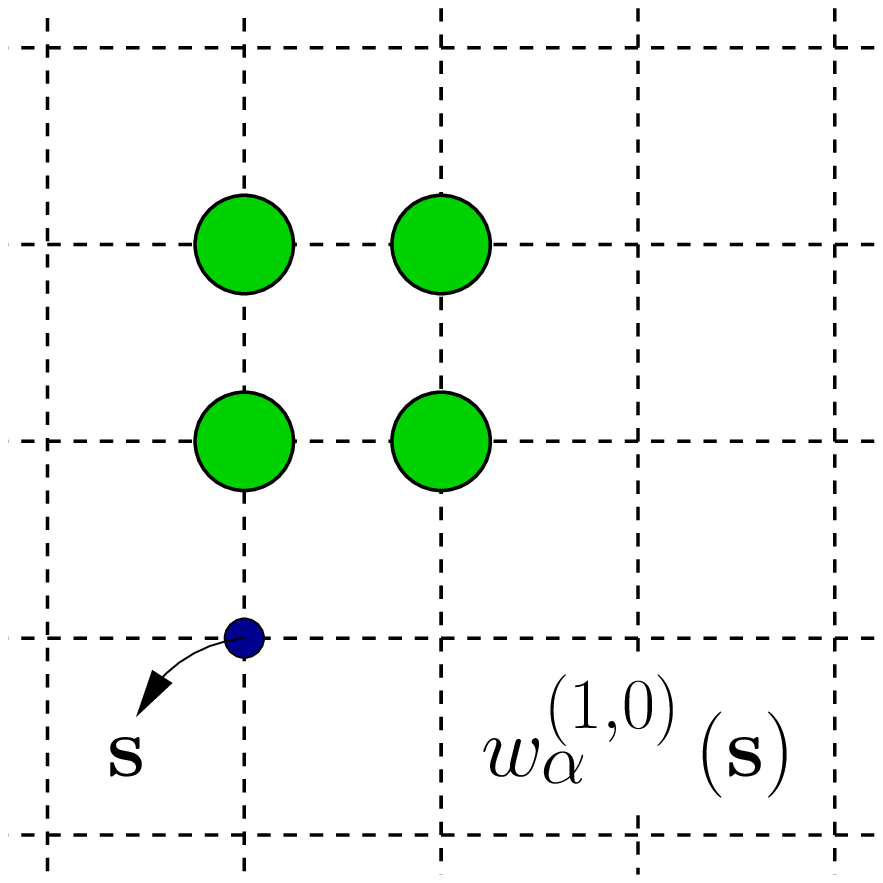}\;\;
  \centering\includegraphics[width=0.23\textwidth]{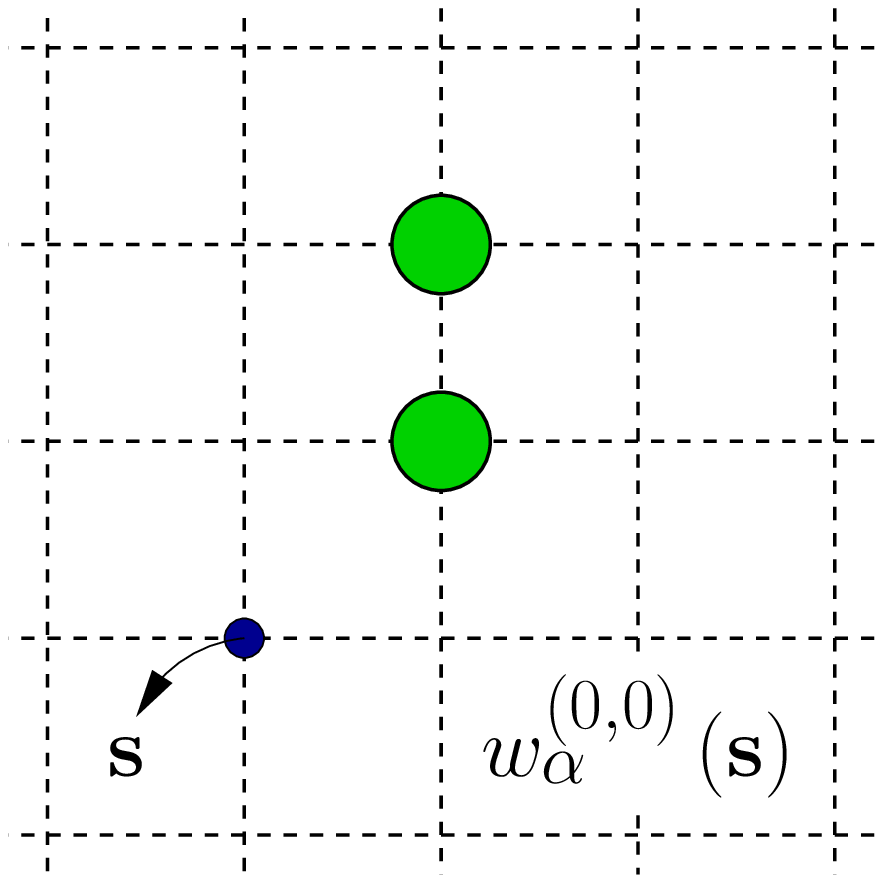}
  \caption{The four FMT weight functions in 2D for a rod with size vector $\vec{L}=\left(2,3\right)$.
           The green circles represents the lattice points at which the weight function is 1.
           The lattice point $\vec{s}$ at which each weight function is evaluated is marked with a blue dot.
           }
  \label{fig:weight}
\end{figure*}

In \crefs{Laf02}{Laf04}
Lafuente and Cuesta have worked out an FMT excess free-energy density
for hard bodies on a lattice model
in the form of \eq{eq:dft_fex_general}.
For a given rod species $\mathcal{L}_\alpha$,
the specified weight functions $w_{i}^\alpha$ and their corresponding weighted densities $n^\alpha$,
are labeled by $d$-dimensional index $\alpha=(\alpha_1,\dots,\alpha_d)$ whose components are either 0 or 1.
Each weight function $w_{i}^\alpha$ can be interpreted as the support of a rod $\mathcal{K}^\alpha_{i}$
whose size vector $\vec{K}^\alpha_{i}=\left(K^{\alpha_1}_{i},\dots, K^{\alpha_d}_{i}\right)$
is related to the size vector  $\vec{L}_i=\left(L_1^i, \dots, L_d^i\right)$ of the original rod species as follows
\bea
  \vec{K}^\alpha_{i} &=& \vec{L}_i - \left(\mathbf{1}_d - \alpha\right)\;,
  \label{eq:weight_supports}
\eea
where $\mathbf{1}_d$ is a $d$-dimensional vector whose components are all 1.
This means the side length of rod $\mathcal{K}^\alpha_{i}$ in dimension $k$ is identical to that of rod $\mathcal{L}_i$
if the corresponding $k$th component of $\alpha$ is 1,
while for $\alpha_k=0$ we have $K^{\alpha_k}_{i}= L^k_i -1$, i.e., it is shortened
by one lattice unit (see \fig{fig:weight}).
In particular, for $\alpha = \mathbf{1}_d$
the size vectors of the support rod $\mathcal{K}^\alpha_{i}$
and the rod species {$\mathcal{L}_i$} are identical.
The corresponding weighted density {$n^\alpha(\vec{s})$}
evaluated at lattice point $\vec{s}$
returns the local packing fraction at that point.

The central physical insight that underlies this choice of weight functions and the subsequent 
construction of the functional 
is dimensional crossover.
By applying an appropriate external potential to the particles of a $d$-dimensional system,
one can restrict the translational degrees of freedom of particles along a given axes
and hence create a system in $(d-1)$ dimensions.
An exact FMT functional should necessarily return the correct excess free energy of such a
dimensionally reduced system.
Of particular interest is the reduction to zero dimensions (0D) by confining the system to
a 0D cavity
which can hold exactly one particle.
The excess free-energy density $\Phi^{\rm 0D}(\eta)$ of a 0D cavity depending on its average occupation $\eta \in[0,1]$ 
is exactly known and reads
\bea
   \Phi^{\rm 0D}\left(\eta\right) &=& \eta + \left(1-\eta\right)\log\left(1-\eta\right)\;.
   \label{eq:dft_phi0D}
\eea
Note that for an $n$-component mixture
one can define a set of 0D cavities $\mathcal{S}_{\rm cav}=\{\mathcal{S}_{\rm cav}^i \}$ with $i=1\dots n$.
The 0D cavity corresponding to species $i$ specifies a minimal set of points  $\mathcal{S}_{\rm cav}^i$ on the lattice
which hold exactly one particle of species $i$.
The local packing fraction in the full (multi-species) cavity
is evaluated as $\eta_{\rm cav}=\sum_{i=1}^n\sum_{\vec{s}\in{\mathcal{S}_{\rm cav}^i}}\rho_i\left(\vec{s}\right)$.
Note that the excess free-energy density of all such 0D cavities is given by \eq{eq:dft_phi0D}
with $\eta=\eta_{\rm cav}$.
Lafuente and Cuesta have derived a functional which returns the correct free energy for all possible 0D cavities
in the system, i.e., it is correct for extreme confinement \cite{Laf02,Laf04}.
The resulting Lafuente-Cuesta excess free-energy density is given by
\bea
  \Phi\left(\{n^{\alpha}\left(\vec{s}\right)\}\right) &=& \mathcal{D}_\alpha \Phi^{\rm 0D}\left(n^{\alpha}\left(\vec{s}\right)\right)\;,
  \quad{\rm with}\;\mathcal{D}_\alpha = \prod_{j=1}^d D_{\alpha_j}\;,
  \label{eq:Phi_LC}
\eea
where $\alpha$ is the $d$-dimensional index as before
and $D_{\alpha_j}$ is a difference operator which acts on a given function $f$ as: $D_{\alpha_j}\;f(\alpha_j)=f(1)-f(0)$.

\subsection{Two dimensions}
For a $\nu$-component mixture of hard rods with short axes of length 1 in 2D,
consider $\nu_{x}$ species parallel to the $x$ axis
and the remaining $\nu_{y}=\nu-\nu_{x}$ oriented along the $y$ axis.
The corresponding excess free-energy density from \eq{eq:Phi_LC} is given by
\bea
   \Phi^{\rm 2D}_{\rm HR}\left(\{n^{\alpha}\}\right) &=& D_{\alpha_1} D_{\alpha_2} \Phi^{\rm 0D}\left(n^{(\alpha_1,\alpha_2)} \right) \nonumber \\
   &=&  \Phi^{\rm 0D}\left(n^{(1,1)}(\vec{s})\right) 
   - \Phi^{\rm 0D}\left(n^{(0,1)}(\vec{s})\right) 
   - \Phi^{\rm 0D}\left(n^{(1,0)}(\vec{s})\right)\;,
   \label{eq:dft_phi2D}
\eea
where the fourth term $\Phi^{\rm 0D}\left(n^{(0,0)}\right)$ vanishes since $n^{(0,0)}=0$ for 
the type of hard rods considered,
and the other weighted densities are calculated as follows:
\bea
   n^{(1,1)}(\vec{s}) &=& \sum_{i=1}^\nu \rho_i \ast w_i^{(1,1)}\;, \nonumber\\
   n^{(0,1)}(\vec{s}) &=& \sum_{i_{x}=1}^{\nu_{x}} \rho_{i_{x}} \ast w_{i_{x}}^{(0,1)}\;,\nonumber\\
   n^{(1,0)}(\vec{s}) &=& \sum_{i_{y}=1}^{\nu_{y}} \rho_{i_{y}} \ast w_{i_{y}}^{(1,0)}\;.
   \label{eq:dft_n2d}
\eea
Here $\rho_i$'s are the density of species $i$, $w_i^{\alpha}$ are their corresponding weight functions,
and $\ast$ denotes the discrete convolution defined in \eq{eq:dft_fex_general}.
Note that for $n^{(0,1)}$ [$n^{(1,0)}$] the sum is restricted to those particles which lie along the $x$ axis [$y$-axis]
since only their corresponding weighted densities are not zero.

For the construction of an FMT-AO functional for attracting rods,
we start from the excess free-energy density of a four-component mixture.
We will consider two colloidal species whose size vectors are given by $\vec{L}_x=(L,1)$ and $\vec{L}_y=(1,L)$.
Moreover, we need two polymer species, $\vec{L}_{{\rm p},x}=(\Lp,1)$ and $\vec{L}_{{\rm p},y}=(1, \Lp)$,  for inducing the attractions.
By setting equal polymer chemical potentials $\mu_{{\rm p}, x}=\mu_{{\rm p}, y}=\mu_{\rm p}$,
and, consequently, equal polymer reservoir densities $\rhopri{x}=\rhopri{y}=\rhopr$,
we ensure a symmetric attractive interaction between colloidal particles along the $x$ and $y$ axes.
The excess free-energy density for the AO model is obtained by linearizing $\Phi^{\rm 2D}_{\rm HR}$ from \eq{eq:dft_phi2D}
with respect to polymer densities [see \eqto{eq:AO_Philin}{eq:AO_conep}].
The required polymeric first-order direct correlation functions $\cone{p}{j}$ are given by
  \bea
     -\cone{p}{x}(\vec s) &=&  
                    - w_{{\rm p},x}^{(1,1)}\hat\ast\log\left(1-n_{\rm c}^{(1,1)}\right) 
                     + w_{{\rm p},x}^{(1,0)}\hat\ast\log\left(1-n_{\rm c}^{(1,0)}\right) 
                  \;,\nonumber \\
     -\cone{p}{y} (\vec s) &=& 
                    - w_{{\rm p},y}^{(1,1)}\hat\ast\log\left(1-n_{\rm c}^{(1,1)}\right) 
                     + w_{{\rm p},y}^{(0,1)}\hat\ast\log\left(1-n_{\rm c}^{(0,1)}\right) 
                  \;.
     \label{eq:2d_conep}
  \eea
Hence, for a given polymer reservoir density $\rhopr$ and after specifying the colloidal density profiles,
the equilibrium polymer densities $\rho_{{\rm p},j}^{\rm eq}$ is obtained using \eqs{eq:ao_rpeq}{eq:2d_conep}.
Consequently, the total free energy $\beta\F_{\rm AO}$ [\eq{eq:ao_F}] and
the semi-grand free energy $\beta\F'_{\rm AO}$ [\eq{eq:ao_Fp}] are obtained.
Finally, the effective free energy for colloidal particles $\beta\F_{\rm AO}^{\rm eff}$
is fully determined as a functional of the colloidal densities $\r{c}{i}$
by using \eq{eq:ao_feff} and the expressions for excluded volumes $V^{\rm excl}_{ij}$ from \eq{eq:ao_vexcl}.

In the following, we explicitly consider sticky attractions and discuss the results for a bulk state.
Therefore, the polymer length which determines the range of attraction is set to $\Lp=2$.
In a bulk (homogeneous) state $\r{c}{x}$ and $\r{c}{y}$ are constant.
As a result, all colloidal weighted densities are constant,
i.e., $n^{(1,1)}=L(\r{c}{x}+\r{c}{y})=L\rhoc=\eta$, where $\rhoc$ is the total colloidal density and $\eta$ is the packing fraction,
$n^{(0,1)}=(L-1)\r{c}{x}$, and $n^{(1,0)}=(L-1)\r{c}{y}$.
Consequently, the corresponding equilibrium polymer densities $\r{p}{j}$ and
polymeric weighted densities are also constant.
Using \eqs{eq:ao_rpeq}{eq:2d_conep}, the equilibrium polymer densities in bulk read
\bea
  \r{p}{j} &=& \rhopr\; e^{\cone{p}{j}} = \rhopr\;\frac{\left(1-L\rhoc\right)^2}{\left(1-(L-1)\r{c}{j}\right) }\qquad{\rm for}\;{j=x,y}\;.
   \label{eq:2d_rhop}
\eea

The total free-energy density of the system $\beta f^{\rm 2D}_{\rm AO}$
can be written as a sum of ideal gas free-energy density of colloidal and polymeric rods,
$\beta f^{\rm id}_{\rm c}$ and $\beta f^{\rm id}_{\rm p}$ respectively,
the entropic contribution to the excess free-energy density $\Phi^{\rm 2D}_{\rm HR}$
and the energetic contribution due to the attractive interactions:
\bea
 \beta f^{\rm 2D}_{\rm AO} &=& \beta f^{\rm id}_{\rm c} +\beta f^{\rm id}_{\rm p}  + \Phi^{\rm 2D}_{\rm HR} {-} \sum_{j=x,y} \r{p}{j}\; \cone{p}{j}\\
 \beta f^{\rm id}_{\rm c} &=&  \sum_{i=x,y} \beta f^{\rm id}\left(\r{c}{i}\right) \\
 \beta f^{\rm id}_{\rm p} &=&  \sum_{j=x,y} \beta f^{\rm id}\left(\r{p}{j}\right)\\
 \Phi^{\rm 2D}_{\rm HR} &=& \Phi^{\rm 0D}\left(L\rhoc\right) - \Phi^{\rm 0D}\left((L-1)\r{c}{x}\right) - \Phi^{\rm 0D}\left((L-1)\r{c}{y}\right)\;,
\eea
where the ideal gas free-energy density $\beta f^{\rm id}(\rho)$ is defined in \eq{eq:dft_ideal}.
Consequently, the semi-grand free-energy density $\beta f'_{\rm 2D}$
and the effective colloidal free-energy density $\beta f^{\rm eff}_{\rm 2D}$ read
\bea
   \beta f'_{\rm 2D}
   &=& \beta f_{\rm 2D}^{\rm AO}
       - \mu_{\rm p} \left( \r{p}{x} + \r{p}{y}\right)\nonumber \\
 &=& \beta f^{\rm id}_{\rm c} + \Phi^{\rm 2D}_{\rm HR} - \left( \r{p}{x} + \r{p}{y}\right)\; \\
   \beta f^{\rm eff}_{\rm 2D}
   &=& \beta f'_{\rm 2D}
   -\rhopr\left(-2+\left(3L+1\right)\rhoc\right) \\
   &=& \beta f^{\rm id}_{\rm c} + \Phi^{\rm 2D}_{\rm HR} - \rhopr \left( \sum_{j=x,y} \frac{\left(1-L\rhoc\right)^2}{\left(1-(L-1)\r{c}{j}\right) }
   + \left(-2+\left(3L+1\right)\rhoc\vphantom{1^1} \right) \right) \;. \nonumber
\eea

The attractive part of the effective free-energy density is linear in $\rhopr$ and the leading term for
small colloidal rod densities is quadratic in these and equivalent to the Bragg-Williams approximation since
the virial expansion of the LC functional is correct up to third order. The Bragg-Williams approximation
of the attractive part gives a term
$ -\rhopr \left[ (L^2+1)(\r{c}{x}^2+\r{c}{y}^2) + 4 L   \r{c}{x} \r{c}{y} \right]$.
The case $L=1$ corresponds to the lattice gas. Setting $\r{c}{x}=\rhoc$ and $\r{c}{y}=0$ [we have only one component
for particles with extension (1,1)], the effective free-energy density reduces to the Bragg-Williams approximation
for all densities:
\bea
  \beta f^{\rm eff}_{\rm 2D} &=& \rhoc \ln \rhoc + (1-\rhoc)\ln(1-\rhoc) - 2\rhopr \rhoc^2 \qquad (L=1)\;.
\eea

\begin{figure*}[!t]
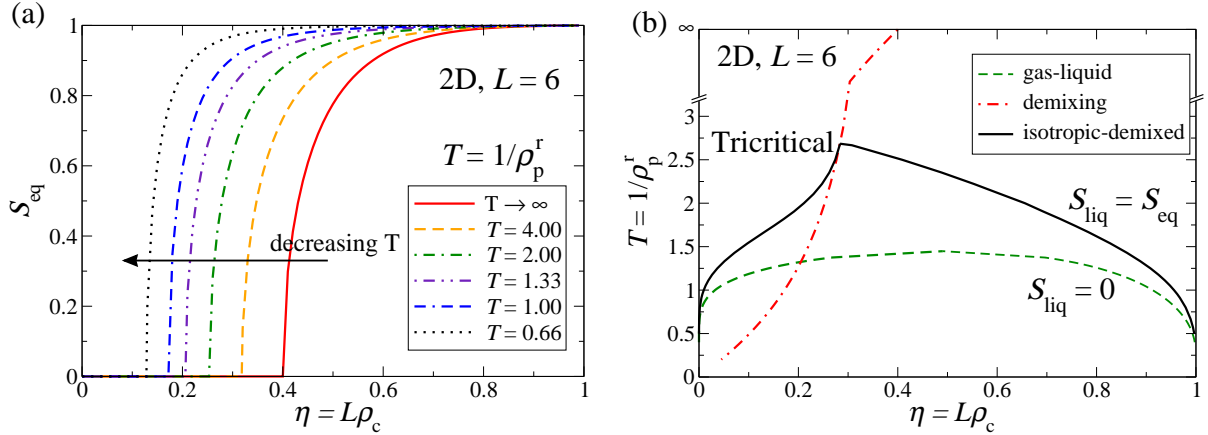

     \centering
     \includegraphics[width=0.47\textwidth]{2d_L6_seq.eps}\;\;\;
     \includegraphics[width=0.47\textwidth]{2d_L6_coex.eps}
     \caption{(a) Equilibrium demixing order parameter for $L=6$ and different polymer reservoir densities.
                 There exists a critical density above which the system demixes.
                 The critical density shifts to lower densities by increasing effective attraction;
                 i.e., decreasing the effective temperature $T=1/\rhopr$ in the AO model.
              (b) The phase diagram of the 2D lattice model for $L=6$.
                 There exists a tricritical point below which the system exhibits a first order phase transition
                 between an isotropic gas phase and a demixed state for which $S_{\rm eq}\neq0$.
                 The isotropic gas-liquid phase transition is unstable with respect to the former phase transition.}
     \label{fig:res_2d}
  \end{figure*}

For the determination of the bulk phase diagram it is useful to introduce an order parameter $S$ for demixing: 
\bea
   S = \frac{\r{c}{x} - \r{c}{y}}{\r{c}{x} + \r{c}{y}}
   \quad\Rightarrow\quad
   \r{c}{x}&=&\frac{\rhoc}{2}\left( 1+S \right)  \;,\nonumber\\
   \r{c}{y}&=&\frac{\rhoc}{2}\left( 1-S \right)  \;.
   \label{eq:2d_order}
\eea
For $L>1$ and for
a given total colloidal density $\rhoc$, the equilibrium value for the demixing order parameter $S_{\rm eq}$ minimizes the effective free-energy density.
For small densities, the mixed state has the minimum free energy, i.e., $S_{\rm eq}=0$ .
On increasing $\rhoc$, we reach a critical density $\rho_{\rm c,cr}$ above which we have a demixed state.
For a pure hard-rod system, it has been shown that for $L\geq 4$ the system demixes at $\rho_{\rm c,cr} = 2/\left[L(L-2)\right]$ \cite{Oet16}.
Taking the attractive interactions into account, this transition shifts to lower densities for a given colloidal rod length $L$.
This is shown in \subfig{fig:res_2d}{a} for rod length $L=6$.

The complete phase diagram for a fixed rod length [$L=6$ in \subfig{fig:res_2d}{b}] reflects the competition between demixing
(present for all $\rhopr$) and the gas-liquid transition (setting in above a critical $\rhopr$).
The gas-liquid binodal for a coexisting isotropic gas state and an assumed isotropic liquid state
is shown by the green dashed line in \subfig{fig:res_2d}{b}. However, the coexisting isotropic liquid state is unstable 
since the liquid branch is above the critical density of demixing [red dot-dashed line in \subfig{fig:res_2d}{b}].
Therefore the gas-liquid binodal [full black line in \subfig{fig:res_2d}{b}] becomes deformed:
The demixing line, starting from the hard-rod limit $\rhopr=0$ ($T=1/\rhopr=\infty$) ends in a tricritical point
below which coexistence between an isotropic gas state and a demixed liquid state is found.

The FMT-AO free-energy density delivers the phase behavior of the competing demixed and gas-liquid phases from a single
expression for the free-energy density. This goes beyond existing theoretical treatments such as in \cref{Lon12} which
have to resort to different free energy models for an isotropic and a fully demixed state. However, a comparison to available
simulation results shows the well-known difficulties of mean-field models in 2D. In \cref{Lon10}, the system is studied
by a mixture of canonical and grand-canonical Monte Carlo methods. It is found that also in the case
of attractions demixing is only present for $L\ge 7$ (as for hard rods). 
Demixing shifts to {\em higher} rod densities
with increasing attractions, in contrast to our theory results here and the results in \cref{Lon12}.
This appears surprising since one might think that the sticky attractions increase the propensity of the rods to align 
and thus favor the demixed phase (with alignment). Control simulations performed by us suggest that the sticky rods 
organize in larger domains of locally aligned rods with no alignment globally. These domains fluctuate strongly in size and shape,
and the entropic contribution of such fluctuations is not captured by the theoretical treatments. 
Furthermore the simulations \tb{of \cref{Lon10}} suggest that the critical point of the gas-(isotropic) liquid transition
survives with increasing attractions. From the simulation results it is not clear whether the demixing line meets the liquid branch
of the gas-liquid binodal in a tricritical point or in a critical end point.

\subsection{Three dimensions}

Consider a $\nu$-component mixture of hard rods with short axes of length 1 in 3D,
of which $\nu_{x}$ species are oriented parallel to the $x$ axis,
$\nu_{y}$ species are oriented parallel to the $y$ axis,
and the remaining $\nu_{z}=\nu-(\nu_{x}-\nu_{y})$ species 
are oriented parallel to the $z$ axis.
The excess free-energy density from \eq{eq:Phi_LC} is in this case
\bea
   \Phi^{\rm 3D}_{\rm HR}\left(\{n^{\alpha}\}\right) &=& D_{\alpha_1} D_{\alpha_2}  D_{\alpha_3} \Phi^{\rm 0D}\left(n^{(\alpha_1,\alpha_2,\alpha_3)} \right) \nonumber \\
    &=&
     \Phi^{\rm 0D}\left(n^{(1,1,1)}(\vec{s})\right) 
   - \Phi^{\rm 0D}\left(n^{(0,1,1)}(\vec{s})\right) 
   - \Phi^{\rm 0D}\left(n^{(1,0,1)}(\vec{s})\right)
   - \Phi^{\rm 0D}\left(n^{(1,1,0)}(\vec{s})\right)\;.
   \label{eq:dft_phi3D}
\eea
Note that all other weighted densities $n^{(\alpha_1,\alpha_2,\alpha_3)}$
with $\alpha_1+\alpha_2+\alpha_3\leq 1$ are zero.
The non-vanishing weighted densities are calculated as follows:
\bea
   n^{(1,1,1)}(\vec{s}) &=& \sum_{i=1}^\nu \rho_i \ast w_i^{(1,1,1)}\;, \nonumber\\
   n^{(0,1,1)}(\vec{s}) &=& \sum_{i_{x}=1}^{\nu_{x}} \rho_{i_{x}} \ast w_{i_{x}}^{(0,1,1)}\;,\nonumber\\
   n^{(1,0,1)}(\vec{s}) &=& \sum_{i_{y}=1}^{\nu_{y}} \rho_{i_{y}} \ast w_{i_{y}}^{(1,0,1)}\;,\nonumber\\
   n^{(1,1,0)}(\vec{s}) &=& \sum_{i_{z}=1}^{\nu_{z}} \rho_{i_{z}} \ast w_{i_{z}}^{(1,1,0)}\;,
   \label{eq:dft_n3d}
\eea
where, as in \eq{eq:dft_n2d}, the sums {in the last three weighted densities} are restricted to the 
species which are extended along the axis {where the corresponding $\alpha$ index is zero}.

For the construction of an FMT-AO functional for attracting 3D rods,
consider three species of colloidal hard rods
with equal length and size vectors $\vec{L}_x=(L,1,1)$, $\vec{L}_y=(1,L,1)$, and
$\vec{L}_z=(1,1,L)$.
For inducing attractions along each axis, we {need  three polymer species with size vectors}
$\vec{L}_{{\rm p},x}=(\Lp,1,1)$, 
$\vec{L}_{{\rm p},y}=(1,\Lp,1)$, 
and $\vec{L}_{{\rm p},z}=(1, 1, \Lp)$.
As in the derivation of the 2D functional, 
we assume that the corresponding polymer reservoir densities $\rhopri{j}$ have the same value $\rhopr$.
Hence, we have to start from the excess free-energy density $\Phi^{\rm 3D}_{\rm HR}$ of a six-component hard-rod mixture in 3D,
i.e., \eq{eq:dft_phi3D} with the weighted densities $n^{\alpha}$ from \eq{eq:dft_n3d}.
Linearization with respect to the polymer densities results in the following polymeric first-order
direct correlation functions: 
  \bea
     -\cone{p}{x} &=&  
                    - w_{{\rm p},x}^{(1,1,1)}\hat\ast\log\left(1-n_{\rm c}^{(1,1,1)}\right) 
                     + w_{{\rm p},x}^{(1,0,0)}\hat\ast\log\left(1-n_{\rm c}^{(1,0,0)}\right) 
                  \;,\nonumber \\
     -\cone{p}{y} &=&  
                    - w_{{\rm p},y}^{(1,1,1)}\hat \ast\log\left(1-n_{\rm c}^{(1,1,1)}\right) 
                     + w_{{\rm p},y}^{(0,1,0)}\hat \ast\log\left(1-n_{\rm c}^{(0,1,0)}\right) 
                  \;,\nonumber \\
     -\cone{p}{z} &=&  
                    - w_{{\rm p},z}^{(1,1,1)}\hat\ast\log\left(1-n_{\rm c}^{(1,1,1)}\right) 
                     + w_{{\rm p},z}^{(0,0,1)}\hat\ast\log\left(1-n_{\rm c}^{(0,0,1)}\right) 
                  \;.
     \label{eq:3d_conep}
  \eea
  Consequently, for a given $\rhopr$ and colloidal rod densities $\r{c}{i}$,
  the equilibrium polymer densities $\rho_{{\rm p},j}^{\rm eq}$,
  the total free energy $\beta\F^{\rm 3D}_{\rm AO}$, the semi-grand free energy $\beta\F'_{\rm 3D}$,
  and, finally, the effective free energy for colloidal {rods only} $\beta\F^{\rm 3D}_{\rm eff}$ are 
  determined by \eqto{eq:ao_F}{eq:ao_vexcl}.

  \begin{figure*}[!t]
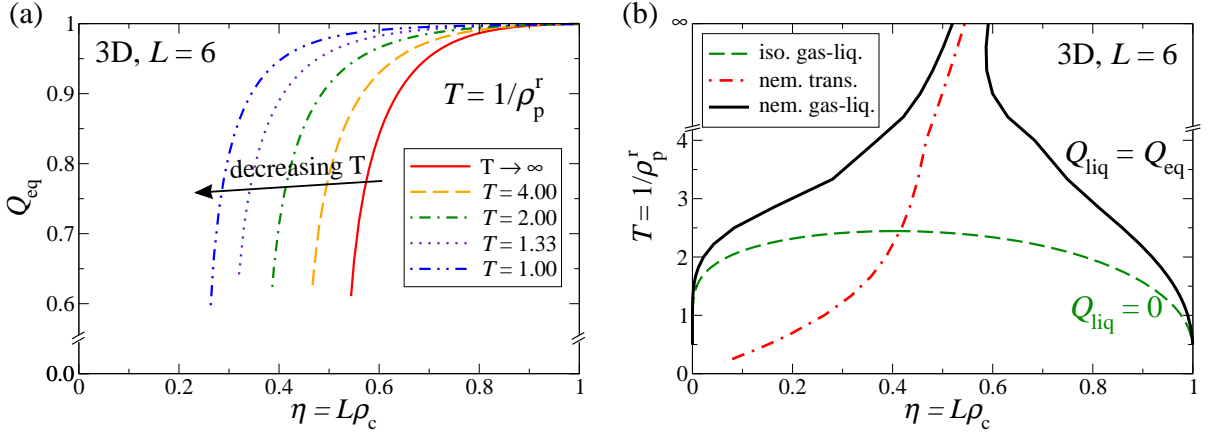

     \centering
     \includegraphics[width=0.47\textwidth]{3d_L6_qeq.eps}\;\;\;
     \includegraphics[width=0.47\textwidth]{3d_L6_coex.eps}
     \caption{(a) Equilibrium nematic order parameter $Q_{\rm eq}$ for $L=6$ and different polymer reservoir densities $\rhopr=1/T$.
                 On increasing the total colloidal density $\rhoc$
                 the system undergoes a first-order isotropic-nematic phase transition.
                 The critical density shifts to lower densities by increasing effective attraction;
                 i.e., decreasing the effective temperature $T=1/\rhopr$ in the AO model.
              (b) The phase diagram of the 3D lattice model for $L=6$.
                 For a pure hard-rod system ($T\rightarrow\infty$),
                 there is a first-order phase transition from an isotropic gas
                 to a nematic liquid state.
                 By decreasing the effective temperature, the phase coexistence region becomes broader.}
     \label{fig:res_3d}
  \end{figure*}

Now we turn to a bulk state where all colloidal densities are constant.
The non-vanishing colloidal weighted densities are given by
$n^{(1,1,1)}=L\left(\r{c}{x}+\r{c}{y}+\r{c}{z}\right)=L\rhoc=\eta$ with $\rhoc$ the total colloidal density and $\eta$ the packing fraction,
$n^{(0,1,1)}=(L-1)\r{c}{x}$, $n^{(1,0,1)}=(L-1)\r{c}{y}$, and $n^{(1,1,0)}=(L-1)\r{c}{z}$.
As in 2D, we will only consider {sticky} attractions, i.e., $\Lp=2$.
Using \eqs{eq:ao_rpeq}{eq:3d_conep}, the equilibrium polymer densities in bulk {become}
\bea
  \r{p}{j} &=& \rhopr\; e^{\cone{p}{j}} = \rhopr\;\frac{\left(1-L\rhoc\right)^2}{\left(1-(L-1)\r{c}{j}\right) }\qquad{\rm for}\;j=x,y,z\;.
   \label{eq:3d_rhop}
\eea

Similar steps as in the 2D case lead to the following effective free-energy density:
\bea
   \beta f^{\rm eff}_{\rm 3D}
   &=& \beta f^{\rm id}_{\rm c} + \Phi^{\rm 3D}_{\rm HR} - \rhopr \left( \sum_{j=x,y,z} \frac{\left(1-L\rhoc\right)^2}{\left(1-(L-1)\r{c}{j}\right) }
   + \left(-3+\left(5L+1\right)\rhoc\vphantom{1^1}\right)  \right) \;.
\eea
with
\bea
   \Phi^{\rm 3D}_{\rm HR} &=& \Phi^{\rm 0D}\left(L\rhoc\right) - \sum_{i=x,y,z} \Phi^{\rm 0D}\left((L-1)\r{c}{i}\right)\;, \\
 \beta f^{\rm id}_{\rm c} &=&  \sum_{i=x,y,z} \beta f^{\rm id}\left(\r{c}{i}\right) \;. \nonumber
\eea
As before, for small $\r{c}{i}$ the attractive part is quadratic in the colloidal rod densities and 
will reduce to the correct form of the Bragg-Williams approximation.

The phase diagram determination is facilitated by the introduction of two
order parameters, usually denoted as the nematic order parameter $Q$
and the biaxiality parameter $S$:
\bea
  Q &=&   \frac{\r{c}{z} - \frac{1}{2}\left(\r{c}{x}+\r{c}{y}\right)}{\r{c}{x} + \r{c}{y} + \r{c}{z}}\;,
  \nonumber \\
  S &=& \frac{\r{c}{x} - \r{c}{y}}{\r{c}{x} + \r{c}{y}}\;,
  \label{eq:3d_order}
\eea
and in terms of these the density of each colloidal species $\r{c}{i}$ is determined as 
\bea
  \r{c}{x} &=&\frac{\rhoc}{3}\left( 1+S \right)\left(1-Q\right)  \;,\nonumber\\
  \r{c}{y} &=&\frac{\rhoc}{3}\left( 1-S \right)\left(1-Q\right)  \;,\nonumber\\
  \r{c}{z} &=&\frac{\rhoc}{3}\left(1+2Q\right)  \;.
  \label{eq:3d_densities}
\eea
For a given colloidal density $\rhoc$,
the equilibrium value of the order parameters, $S_{\rm eq}$ and $Q_{\rm eq}$,
is obtained by a simultaneous minimization of the effective free-energy density.
It turns out that for a given polymer reservoir density,
$S_{\rm eq}=0$ for all densities.
However, the nematic order parameter $Q$ shows a first-order transition
from an isotropic state $Q_{\rm eq}=0$ to a nematic state with $Q_{\rm eq}>0$.
For hard rods, the transition is present for $L \ge 4$ and the associated
critical packing fraction $\eta_{\rm cr}(L)$
shifts to lower packing fractions for longer rods \cite{Oet16}. Coexisting isotropic and
nematic states are separated by a substantial density gap.
On switching on the attractions, we find, generally, that the critical packing fraction $\eta_{\rm cr}(L,\rhopr)$
decreases on increasing $\rhopr$ and that the density gap between coexisting isotropic and nematic states
continuously widens (for $L={\rm const.}$ and $L \ge 4$). 
We illustrate this for $L=6$ in \fig{fig:res_3d}.
Figure \ref{fig:res_3d}(a)
shows the behavior of $Q_{\rm eq}$ for different effective temperatures $T = 1/\rhopr$ which displays
the discontinuous jump as well as the shift of the critical packing fraction to lower values
on decreasing $T$.
Figure \ref{fig:res_3d}(b) shows the corresponding phase diagram.
As in the 2D case, we have calculated the binodal for an isotropic gas-liquid transition ($S=Q=0$, green dashed line).
The line of critical packing fractions $\eta_{\rm cr}(L=6,\rhopr=1/T)$ is shown by the red dot-dashed line and
one sees that the liquid branch of the isotropic gas-liquid  binodal 
is unstable with respect to the onset of nematic order. Therefore the physical binodal (full black line) corresponds
to coexistence between an isotropic, lower density state and a nematic, higher-density state for all $T=1/\rhopr$.
The density gap is continuously increasing with decreasing $T$ and thus the isotropic-nematic transition
smoothly acquires the character of a gas-liquid transition as well.

Simulation results for 3D lattice rods with attractions are not available, whereas for rods in the continuum there are 
\cite{Bol97}. For continuum rods at higher densities, there is a transition from the nematic phase to a smectic and a crystalline
phase. On increasing attractions, the isotropic-nematic transition becomes unstable in favor of the more ordered
smectic and crystalline phases. Therefore one observes a similar widening of the coexistence gap on increasing the attractions
as in the lattice model but the coexisting states correspond to an isotropic gas state and either a solid state
(when polymers induce the attractions) or a smectic state (when there is an explicit pairwise, attractive potential between the rods) 
\cite{Bol97}.


\subsection{Monolayer ((2+1)D)}

Here we consider a monolayer of rods on a substrate which can lie down or stand up. It can serve as a toy model for
Langmuir monolayers or a thin film of anisotropic organic molecules as often investigated in the context of research
on organic semiconductors \cite{review-OMBD-Schreiber,review-OMBD-Witte-Woell}.
Effectively, the system can be mapped onto a mixture of $\nu$ species in 2D.
The $\nu_{z}$ species of standing-up rods are treated as their projection on the substrate,
i.e. particles with size vector $\vec{L}_{i_{z}}=(1,1)$.
The remaining $\nu_{x}+\nu_{y}$ species are defined as in the 2D system.
The corresponding excess free-energy density $\Phi^{\rm (2+1)D}_{\rm HR}$ is the same as in \eq{eq:dft_phi2D}
with the weighted densities provided by \eq{eq:dft_n2d}.
Note that in calculating $n^{(1,1)}$ the corresponding weighted density of
``standing-up'' rods, $\rho_{i_z}\ast w_{i_z}^{(1,1)}$, are also considered.

\begin{figure}[!t]
  \centering
  \includegraphics[width=0.23\textwidth]{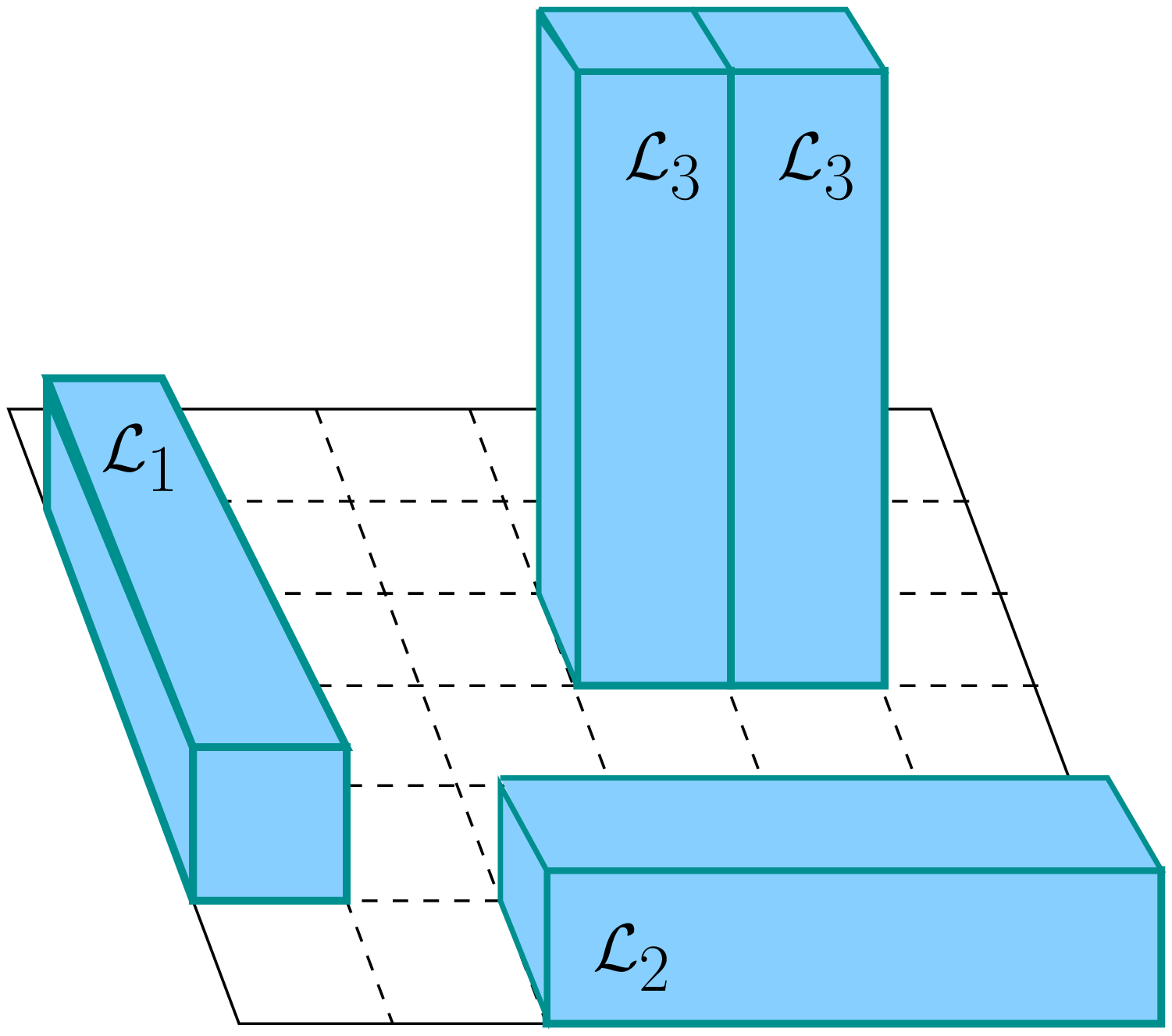}\;\;
  \includegraphics[width=0.23\textwidth]{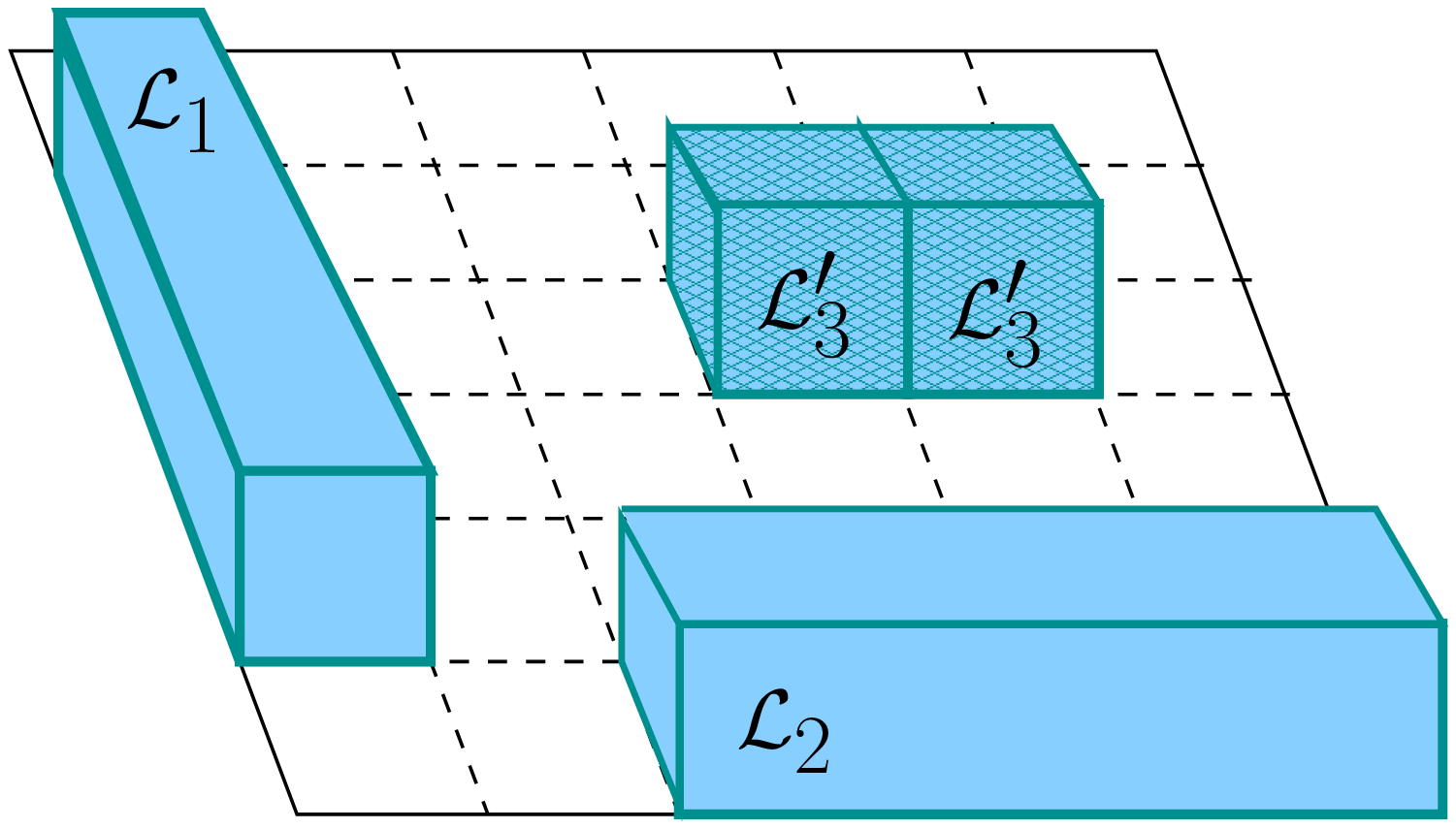}
  \caption{A 3D monolayer in which the particles are confined to move on a substrate (left)
           can be translated to a 2D system where the projection of standing-up rods is considered as a new species
           with size $(1,1)$ (right).}
  \label{fig:2p1d}
\end{figure}

For constructing an FMT-AO functional for sticky rods in (2+1)D,
consider three species of colloidal rods,
with size vectors denoted by $\vec{L}_x=(L,1)$ and $\vec{L}_y=(1,L)$ for lying-down rods
and  $\vec{L}_z=(1,1)$ for standing-up rods.
By adding two polymer species, $\vec{L}_{{\rm p},x}=(\Lp,1)$ and $\vec{L}_{{\rm p},y}=(1, \Lp)$ 
and with reservoir density $\rhopr$ each,
the in-plane attractive interactions are ensured.
However, the out-of-plane interactions of two neighboring standing-up rods is underestimated by a factor of $(L-1)\rhopr$ (see \fig{fig:2p1d}).
In order to compensate this, we will add two more polymer species
$\vec{L}_{{\rm p},xz}=(\Lp,1)$ and $\vec{L}_{{\rm p},yz}=(1, \Lp)$
which only interact with the standing-up rods and have an enhanced polymer reservoir density $\rhopri{xz}=\rhopri{yz}=(L-1)\rhopr$.
As a result, we are dealing with the excess free-energy density of a 2D hard-rod mixture $\Phi^{\rm 2D}_{\rm HR}$ [\eq{eq:dft_phi2D}],
with seven components:
three colloidal and four polymeric species.
Moreover, in linearization of $\Phi^{\rm 2D}_{\rm HR}$ with respect to polymer densities,
there is a slight difference to the 2D case:
the additional polymer species, $\r{p}{xz}$ and $\r{p}{yz}$,
are not interacting with in-plane colloidal rods, $\r{c}{x}$ and $\r{c}{y}$.
As a result,
in calculation of corresponding $\cone{p}{j}$'s,
the density of $\r{c}{x}$ and $\r{c}{y}$ should be set to zero as well.
$\cone{p}{x}$ and $\cone{p}{y}$ are determined similar to those of 2D case.
The polymeric first-order direct correlation functions in this (2+1)D system are given as follows:
  \bea
     -\cone{p}{x} &=&
                    - w_{{\rm p},x}^{(1,1)}\hat\ast\log\left(1-n_{\rm c}^{(1,1)}\right) 
                     + w_{{\rm p},x}^{(1,0)}\hat\ast\log\left(1-n_{\rm c}^{(1,0)}\right) 
                  \;,\nonumber \\
     -\cone{p}{y} &=&
                    - w_{{\rm p},y}^{(1,1)}\hat\ast\log\left(1-n_{\rm c}^{(1,1)}\right) 
                     + w_{{\rm p},y}^{(0,1)}\hat\ast\log\left(1-n_{\rm c}^{(0,1)}\right) 
                  \;,\nonumber\\
     -\cone{p}{xz} &=&
                    - w_{{\rm p},xz}^{(1,1)}\hat\ast\log\left(1-n_{{\rm c},z}^{(1,1)}\right) 
                  \;,\nonumber \\
     -\cone{p}{yz} &=&
                    - w_{{\rm p},yz}^{(1,1)}\hat\ast\log\left(1-n_{{\rm c},z}^{(1,1)}\right) 
                  \;.
     \label{eq:2p1d_conep}
  \eea

After fixing $\rhopr$, the equilibrium density of polymer species $\rho_{{\rm p},j}^{\rm eq}$
are determined by \eq{eq:ao_rpeq} with $\cone{p}{j}$ from {\eq{eq:2p1d_conep}}.
Consequently, the total free energy $\beta\F^{\rm (2+1)D}_{\rm AO}$
and the semi-grand free energy $\beta\F'_{\rm (2+1)D}$ are determined from \eqs{eq:ao_F}{eq:ao_Fp}.
Finally by using \eqs{eq:ao_feff}{eq:ao_vexcl} we obtain an effective free energy for colloidal particles $\beta\F^{\rm (2+1)D}_{\rm eff}$
for a (2+1)D system.

  For a bulk state, the density of colloidal rods $\r{c}{i}$
  and consequently their corresponding weighted densities $n_{\rm c}^{\alpha}$ are constant,
  $n^{(1,1)} = L\left(\r{c}{x} + \r{c}{y}\right) + \r{c}{z}=\eta$,
  $n^{(0,1)} = (L-1)\r{c}{x}$, and
  $n^{(1,0)} = (L-1)\r{c}{y}$.
  For sticky attractions  $\Lp=2$,
  the equilibrium density of polymeric rods
  is obtained by combining \eqs{eq:ao_rpeq}{eq:2p1d_conep}.
  \bea
    \r{p}{j} &=& \rhopr\; e^{-\cone{p}{j}} = \rhopr\;\frac{\left(1-\eta\right)^2}{\left(1-(L-1)\r{c}{j}\right) }\qquad{\rm for}\;j=x,y\;,\nonumber\\
    \r{p}{j} &=& \left(L-1\right) \rhopr\; e^{-\cone{p}{j}} = \rhopr\; \left(L-1\right)\;      \left(1-\r{c}{z}\right)^2\qquad{\rm for}\;j=xz,yz\;.
     \label{eq:2p1d_rhop}
  \eea
Similar steps as in the two cases before lead to the following effective free-energy density:
\bea
   \beta f^{\rm eff}_{\rm (2+1)D}
   &=& \beta f^{\rm id}_{\rm c} + \Phi^{\rm (2+1)D}_{\rm HR}
   -\rhopr\left( \sum_{j=x,y} \frac{\left(1-\eta\right)^2}{\left(1-(L-1)\r{c}{j}\right) } + 2\left(L-1\right)\;\left(1-\r{c}{z}\right)^2 + \right. \nonumber \\
    & & \left. \phantom{\sum_j=x} \left( -2L + \left( 3L+1 \right)\left(\r{c}{x}+\r{c}{y}\right) + 4L \r{c}{z}\vphantom{1^1}\right) \right)\;.
\eea
with
\bea
   \Phi^{\rm (2+1)D}_{\rm HR} &=& \Phi^{\rm 0D}\left( \eta \right) - \sum_{i=x,y} \Phi^{\rm 0D}\left((L-1)\r{c}{i}\right)\;, \\
 \beta f^{\rm id}_{\rm c} &=&  \sum_{i=x,y,z} \beta f^{\rm id}\left(\r{c}{i}\right) \;. \nonumber
\eea
Here the attractions between only the standing rods is equivalent to the Bragg-Williams approximation, whereas
for all other attractions corrections to the Bragg-Williams approximation are present for higher colloidal densities.

The determination of phase diagrams proceeds via the introduction of order parameters as in the 3D case [see \eqs{eq:3d_order}{eq:3d_densities}].
Equilibrium value of the demixing $S_{\rm eq}$ and the nematic $Q_{\rm eq}$ order parameters
are obtained by minimizing the effective free-energy density $\beta f^{\rm eff}_{\rm (2+1)D}$
with respect to corresponding order parameters.
This implies that for a given rod length $L$ and temperature $T=1/\rhopr$,
the corresponding chemical potentials
$\mu_{\rm S}(\rhoc, S,Q)= \partial f^{\rm eff}_{\rm (2+1)D}/\partial S$
and 
$\mu_{\rm Q}(\rhoc, S,Q)= \partial f^{\rm eff}_{\rm (2+1)D}/\partial Q$
should be zero.

  \begin{figure*}[!t]
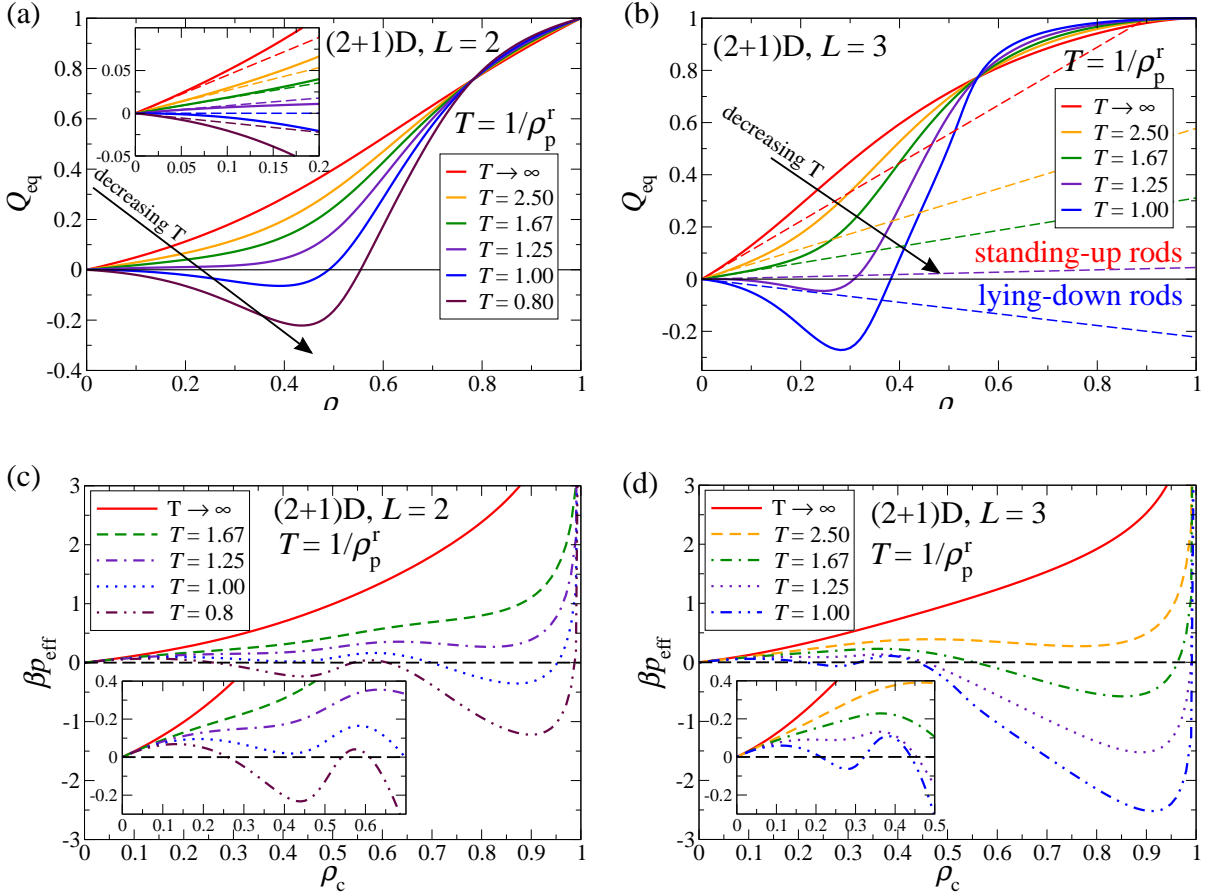

     \centering
     \includegraphics[width=0.47\textwidth]{2p1d_L2_qeq.eps}\;\;\;
     \includegraphics[width=0.47\textwidth]{2p1d_L3_qeq.eps} \\ \vspace*{5mm}
     \includegraphics[width=0.47\textwidth]{2p1d_L2_peff.eps}\;\;\;
     \includegraphics[width=0.47\textwidth]{2p1d_L3_peff.eps}
     \caption{Equilibrium nematic order parameter and resulting surface pressure for a 
              (2+1)D system with $L=2$ [(a) and (c)] and $L=3$ [(b) and (d)].
              Panels (a) and (b): $Q_{\rm eq}$ vs. $\rho_{\rm c}$ (total density) for various effective temperatures
              $T=1/\rhopr$ (full lines). Dashed lines correspond to the low-density expansion [see \eq{eq:qeq_2p1d}].
              Panels (c) and (d): Surface pressure $\beta p = \rho_{\rm c}^2\; \partial/\partial \rhoc\left(f/\rhoc\right)$ as a function of $\rhoc$
              for various $T$. Insets show the onset of the second van~der~Waals loop for temperatures near
              $T_{\rm cr, 2} \approx 1.2$.}
     \label{fig:res_2p1d_eq}
  \end{figure*}

For a pure hard-rod system and in the regime of small rods $L \leq 12$,
$S_{\rm eq}=0$ for all densities and 
a continuous transition from an isotropic state $Q_{\rm eq}=0$ at $\rhoc=0$
to a nematic state $Q_{\rm eq}=1$ for a fully packed system is observed.
For larger rods a reentrant demixing occurs for a certain interval $\rhoc\in\left[\rho_{\rm c, low}(L):\rho_{\rm c, high}(L)\right]$
\cite{Oet16}.
With attractions, such a reentrant behavior persists and occurs also for lower $L$,
which is understandable since
for an attractive system in 2D and for a given rod length $L$,
the demixing transition density shifts to lower values
on decreasing the effective temperature $T=1/\rhopr$ (see \fig{fig:res_2d}).
Hence, for a monolayer one expects to find a certain temperature below which
the planar rods are demixed for $\rhoc \geq \rho_{\rm c, low}(T; L)$.
Since the rods eventually stand up with increasing total density,
there exists a higher density $\rhoc \geq \rho_{\rm c, high}(T; L)$ at which {the lying} rods mix again and $S_{\rm eq}=0$.
In order to calculate $\rho_{\rm c, low}(T; L)$ and $\rho_{\rm c, high}(T; L)$,
we start from obtaining $Q_{\rm eq}(\rhoc)$ by setting $\mu_{\rm Q}(\rhoc, S=0, Q=Q_{\rm eq}) = 0$.
Expanding $\mu_{\rm S}$,
\bea
   \mu_{\rm S}(\rhoc, S, Q) = \mu_{1, S}(\rhoc, Q) S + \mu_{ 3, S}(\rhoc, Q) {S^3} + \cdots\;,
\eea
the de- and remixing densities are obtained numerically by computing the densities at which $\mu_{1, S}(\rho_{\rm c}, Q_{\rm eq})=0$.
In general, we find that for moderate attractions $T=1/\rhopr \gtrsim 1$ demixing is relevant for $L\ge 4$ and that the 
phase diagram becomes very complicated due to the competition of upright (nematic) ordering, demixing in the plane
and the gas-liquid transition. However, the comparison to available simulation results in 2D has shown that
FMT-AO overestimates the tendency to demix in the substrate plane. Therefore we focus on shorter rods ($L=2$ and 3) for the
calculation of equilibrium order and phase transitions.

The equilibrium value of the nematic ordering parameter $Q_{\rm eq}$ for $L=2$ and $L=3$ is shown in \subfigs{fig:res_2p1d_eq}{a}{b}.
Nematic order, i.e., $Q_{\rm eq} \neq 0$, sets in already at $\rhoc=0$. In order to obtain
the low-density behavior, we expand $\mu_Q$ (assuming $S=0$):
\bea
   \mu_{\rm Q}
    = \pder{f_{\rm eff}^{\rm (2+1)D}}{Q}
   &\approx& \frac{2}{3}\rho_{\rm c}\ln\frac{1+2Q}{1-Q} - \frac{2}{9}\rho_{\rm c}^2 \left[\left(L^2+L-2\right)-\left(L-1\right)^2Q\right]
 \nonumber \\
   && -\frac{4}{9}\rho_{\rm c}^2\;\rhopr \; (L-1)\left[L-(L-5)Q\right] + \mathcal{O}\left(\rho_{\rm c}^3\right) \;.
\eea
The equilibrium nematic order parameter is obtained by setting $\mu_Q=0$, {and in leading order in $\rhoc$ it is given by}
\bea
   Q_{\rm eq} &\approx& \frac{1}{9}\rhoc\left(L-1\right)\left[L+2 -  2 L\rhopr   \right]\;.
  \label{eq:qeq_2p1d}
\eea
The nematic order parameter is linear in the total rod density [shown by the dashed lines in \subfigs{fig:res_2p1d_eq}{a}{b}].
Starting from a pure hard-rod system ($T=1/\rhopr\rightarrow\infty$) for a given rod length,
the tendency to order upright becomes weaker as the temperature is decreased.
Eventually, one reaches a certain temperature $T_{{\rm cr}, Q} = 2 L/(L+2)$
below which $Q_{\rm eq}<0$ and the rods preferably order in-plane for small densities.

\begin{figure*}[!t]
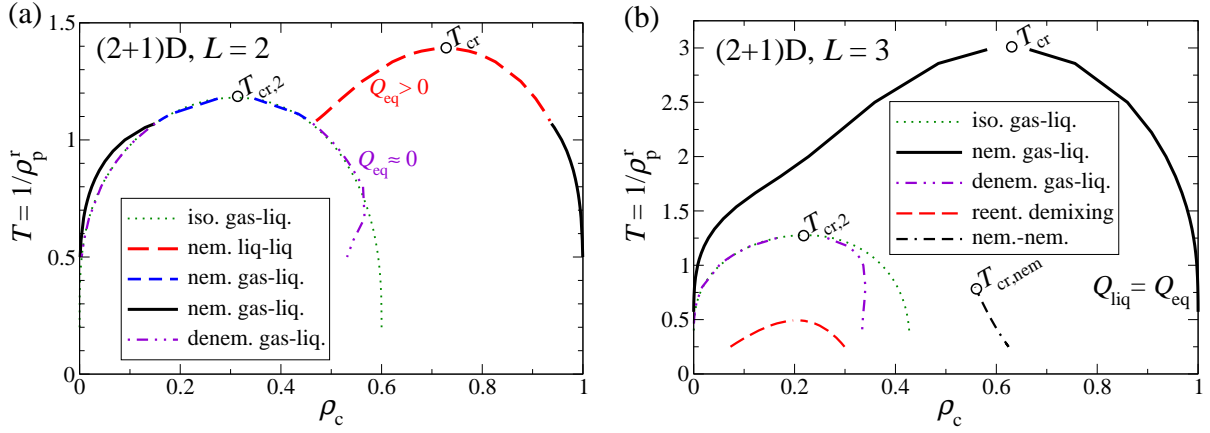

     \centering
     \includegraphics[width=0.47\textwidth]{2p1d_L2_coex.eps}\;\;\;
     \includegraphics[width=0.47\textwidth]{2p1d_L3_coex.eps}
     \caption{Phase diagram of a (2+1)D system for (a) $L=2$ and (b) $L=3$.
              For $L=2$ two stable phase transitions occur for $T<T_{\rm cr}$ {(red dashed binodal) } and $T<T_{\rm cr, 2}$ 
              (blue dashed binodal).
              These transitions become metastable below a triple temperature $T_{\rm tr}$  
              with respect to a phase transition between a highly ordered state at high densities and a 
              nearly isotropic gas state (full black binodal). The purple dot-dot-dashed binodal is the
              metastable continuation of the second binodal.
              For $L=3$, the first transition for $T<T_{\rm cr}$ is stable (full black binodal). The second
              transition for $T<T_{\rm cr, 2}$ (purple dot-dot-dashed binodal) is completely metastable.
              Reentrant demixing in the substrate plane (red dashed line) and
              a discontinuous jump in $Q_{\rm eq}$ (black dash-dash-dotted line) are metastable as well.
              For both $L=2$ and 3 the green dotted line is the binodal of a gas-liquid transition between isotropic states.
             } 
     \label{fig:res_2p1d_phase}
  \end{figure*}

In \subfigs{fig:res_2p1d_eq}{c}{d} we show results for the pressure for $L=2$ and 3, respectively. 
For effective temperatures below an upper critical temperature, $T<T_{\rm cr}$, the van~der~Waals loop
points to a stable phase coexistence between
a low-ordered state at a smaller density and an upright-ordered state at a larger density.
By further decreasing the temperature, a secondary loop is observed whose interpretation will be different for $L=2$ and
$L=3$. The associated phase diagrams for $L=2$ and 3 are shown in \fig{fig:res_2p1d_phase} which are shown
in the plane with axes total colloidal density $\rhoc$ and effective temperature $1/\rhopr$. For $L=3$
[\subfig{fig:res_2p1d_phase}{b}] the upper critical temperature is at $T\approx 3$ and the binodal for the coexisting
states (low density and low order vs. higher density and upright order) is stable for all temperatures
(full black line). For $L=2$ [\subfig{fig:res_2p1d_phase}{a}] the upper critical temperature is at $T\approx 1.4$
and the corresponding binodal is shown by the red dashed line. Below a second, lower critical temperature
$T_{\rm cr, 2} \approx 1.2$ a second stable binodal appears (blue dashed line) which marks coexistence between
two low-ordered states. This second binodal is actually akin to the gas-liquid transition between isotropic states
which we have computed by setting $Q=S=0$ (green dotted line) and which is very close to the second binodal. 
Both the first and second binodals become unstable below a triple temperature $T_{\rm tr} \approx 1.1$
and give way to a binodal marking the coexistence between a nearly isotropic gas and a
highly ordered liquid at high densities (full black line). 
For $L=3$ [\subfig{fig:res_2p1d_phase}{b}] the second binodal is inside the first one ($T_{\rm cr, 2} \approx 1.2$) 
and thus metastable (purple dot-dot-dashed line).
Again, it is almost on top of the gas-liquid binodal for isotropic states (green dotted line).    
For $L=3$, there are two more features in the phase diagram. The red dashed line shows the region of reentrant demixing in the
substrate plane which occurs at low temperatures $T \lesssim 0.5$. The black dash-dash-dotted line corresponds to a discontinuous
jump in the nematic order parameter $Q_{\rm eq}$ which sets in at a third critical temperature $T_{\rm cr, nem}$ and which
would give rise to a first-order nematic-nematic transition. Both transitions (reentrant demixing and nematic-nematic) are
metastable for $L=3$ but would become stable for higher $L$ according to FMT-AO.

These results suggest that the phase diagrams of monolayers can be extremely rich. Previous studies have identified the 
transition between a nearly isotropic gas state and a high density, upright-ordered state \cite{Boehm77,Kra92}, corresponding to
the full black line in \subfig{fig:res_2p1d_phase}{b}. This should be the stable transition for intermediate $L$ (\cref{Kra92} 
confirms this also
by performing Monte Carlo simulations for $L=4$). We emphasize that this is {\em not} the ``descendant'' of the gas-liquid transition
between isotropic states but rather a new nematic liquid-liquid transition. It would be very interesting to check with simulations whether 
the two critical points associated with this ``new'' nematic liquid-liquid transition and the ``old'' gas-liquid transition   
are stable for $L=2$, as we have found here. Such investigations should also be extended to monolayers in the continuum with
short rods. We have not explored a possible substrate potential as an additional degree of freedom which in our opinion may shift the
onset of metastability for the various transitions quite substantially.


\section{Summary and outlook}
\label{sec:summary}

In this work, we have derived a density functional for a lattice model with attractive anisotropic particles (rods). The attractions
are induced by lattice polymers which interact hard with the rods and are an ideal gas amongst themselves (Asakura-Oosawa model).
The functional is derived from a multi-component hard rod functional (for rods and polymers) via linearization with respect 
to the polymeric components. Explicit functionals are obtained by using the Lafuente-Cuesta functional \cite{Laf02,Laf04} for
the multi-component hard rod system. We have applied the functional to the calculation of phase diagrams for sticky rods of length $L$ 
in 2D, 3D, and in a monolayer system [(2+1)D]. In all cases, there is a competition between ordering and gas-liquid transitions.
In 2D, this gives rise to a tricritical point, whereas in 3D, the isotropic-nematic transition crosses over smoothly to a
gas-nematic liquid transition. The richest phase behavior is found for the monolayer system on a neutral substrate. 
For $L=2$, we find two stable
critical points corresponding to the isotropic gas-liquid transition and a nematic liquid-liquid transition. For $L=3$,
the isotropic gas-liquid transition becomes metastable. There are further metastable transitions such as reentrant demixing
in the substrate plane and a nematic-nematic first-order transition. These become stable for larger $L$ but we have not investigated 
this in detail. 

In this work we have not exploited yet the capabilities of our explicit functional in investigating inhomogeneous situations
(correlation functions, interfaces between coexisting states, or wetting/surface transitions on substrates). This will be done in
future work. Of particular interest is also the description of film growth on substrate via a suitable lattice dynamic density
functional theory. First steps in this direction have been taken by calculating the growth of a hard rod monolayer \cite{Klo17}
which shows satisfactory agreement between dynamic DFT and kinetic Monte Carlo simulations. The extension of these investigations
to attractive rods is desired to connect better to actual experimental systems.

{\bf Acknowledgment:} 
  This work is supported within the DFG/FNR INTER project 
  ``Anisotropic Thin Film Growth" by the Deutsche Forschungsgemeinschaft (DFG), Project No. OE 285/3-1.



\begin{thebibliography}{99}

  \bibitem{BookHuang} K. Huang, {\em Statistical Mechanics} (Wiley, New York, 1987), Chap. 14.

%
%
  \bibitem{Bas13} V. R. Dugyala, S. V. Daware, and M. G. Basavaraj
    Soft Matter {\bf 9}, 6711 (2013).

  \bibitem{Sac11} S. Sacanna and D. J. Pine,
    Current Opinion in Colloid and Interface Science {\bf 16}, 96 (2011).

%
%
\bibitem{DiMar61} E. A. DiMarzio,
        J. Chem. Phys. {\bf 35}, 658 (1961).

\bibitem{Alb71} R. Alben,
        Mol. Cryst. and Liq. Cryst. {\bf 13}, 193 (1971).

  \bibitem{Oet16} M. Oettel, M. Klopotek, M. Dixit, E. Empting, T. Schilling, and H. Hansen–Goos,
    J.~Chem.~Phys. {\bf 145}, 074902 (2016).

\bibitem{Cot69} M. A. Cotter and D. E. Martire,
        Mol. Cryst. and Liq. Cryst. {\bf 7}, 295 (1969).

\bibitem{Dhar11} D. Dhar, R. Rajesh, and J. F. Stilck,
        Phys. Rev. E {\bf 84}, 011140 (2011).

\bibitem{Ghosh07} A. Ghosh and D. Dhar,
         Europhys. Lett. {\bf 78}, 20003 (2007).

\bibitem{Gschwind17} A. Gschwind, M. Klopotek, Y. Ai, and M. Oettel,
         Phys. Rev. E {\bf 96}, 20003 (2017).

\bibitem{Raj17} N. Vigneshwar, D. Dhar, and R. Rajesh,
{\em  Different phases of a system of hard rods on three dimensional cubic lattice},
        arXiv:1705.10531.

\bibitem{BookHill} T. L. Hill, {\em An Introduction to Statistical Thermodynamics}
(Addison-Wesley, Reading, MA, 1960), Chap. 14.

\bibitem{Lon12} P. Longone, M. Davila, and A. J. Ramirez-Pastor,
    Phys. Rev. E {\bf 85}, 011136 (2012).

\bibitem{Lon10} P. Longone, D. H. Linares, and A. J. Ramirez-Pastor,
    J.~Chem.~Phys. {\bf 132}, 184701 (2010).

\bibitem{Boehm77} R. E. Boehm and D. E. Martire,
    J.~Chem.~Phys. {\bf 67}, 1061 (1977).

\bibitem{Kra92} D. Kramer, A. Ben-Shaul, Z.-Y. Chen, and W. M. Gelbart,
         J. Chem. Phys. {\bf 96}, 2236 (1992).

%
%
  \bibitem{AO54} S.~Asakura and F.~Oosawa,
    J.~Chem.~Phys. {\bf 22}, 1255 (1954) and
    J.~Polym.~Sci. {\bf 33}, 183 (1958).

  \bibitem{Vri76} A.~Vrij,
    Pure~Appl.~Chem. {\bf 48}, 471 (1976).

  \bibitem{Schm00} M. Schmidt, H. L\"owen, J. M. Brader, and R.~Evans,
    Phys. Rev. Lett. {\bf 85}, 1934 (2000).

\bibitem{Bra03} J. M. Brader, R. Evans, and M. Schmidt,
    Mol. Phys. {\bf 101}, 3349 (2003).

  \bibitem{Laf02} L.~Lafuente and J.~A.~Cuesta,
    J.~Phys.: Condens.~Matter {\bf 14}, 12079 (2002).

  \bibitem{Laf04} L.~Lafuente and J.~A.~Cuesta,
    Phys.~Rev.~Lett. {\bf 93}, 130603 (2004).

\bibitem{Mor16} M. Mortazavifar and M. Oettel,
     J.~Phys.: Condens.~Matter {\bf 28}, 244018 (2016).

  \bibitem{Eva79} R.~Evans,
    Advances~in~Physics {\bf 28}, 143 (1979).

\bibitem{Rot10} R.~Roth,
        J.~Phys.: Condens.~Matter {\bf 22}, 063102 (2010).

\bibitem{Arch17} A. J. Archer, B. Chacko, and R. Evans,
       J.~Chem.~Phys.  {\bf 147}, 034501 (2017).

  \bibitem{San15} A. Santos, M. Lopez de Haro, G. Fiumara, and F. Saija,
    J.~Chem.~Phys {\bf  142}, 224903 (2015).

 \bibitem{BraderThesis} J.~M.~Brader,
     {\em Statistical Mechanics of a Model Colloid-Polymer Mixture},
     Ph.D dissertation, H. H. Wills Physics Laboratory, University of Bristol (2001),
     \url{http://www.bristol.ac.uk/physics/media/theory-theses/brader-jm-thesis.pdf}.

 \bibitem{Schmidt05} J.~A.~Cuesta, L.~Lafuente, and M.~Schmidt,
    Phys.~Rev.~E {\bf 72}, 031405 (2005).


\bibitem{Bol97} P. G. Bolhuis, A. Stroobants, D. Frenkel, and H. N. W. Lekkerkerker,
    J.~Chem.~Phys. {\bf 107}, 1551 (1997).


\bibitem{review-OMBD-Schreiber} F. Schreiber,
         Phys. Stat. Sol. A {\bf 201}, 1037 (2004).

\bibitem{review-OMBD-Witte-Woell} G. Witte and C. W{\"o}ll,
         J. Materials Res. {\bf 19}, 1889 (2004).


\bibitem{Klo17} M. Klopotek, H. Hansen-Goos, M. Dixit, T. Schilling, F. Schreiber, and M. Oettel,
          J.~Chem.~Phys {\bf 146}, 084903 (2017).
 


  



\end{thebibliography}
\end{document}